\documentclass[aps, longbibliography,twocolumn,showpacs,amsmath,amssymb,superscriptaddress,pra]{revtex4-1}
\usepackage[utf8]{inputenc}

\usepackage{tikz}
\usepackage{graphicx}
\usepackage{amsfonts}
\usepackage{amsmath}
\usepackage{amssymb,bbold}
\usepackage{color}
\usepackage{bm}
\usepackage{bbm}
\usepackage{dsfont}
\usepackage{enumerate}
\usepackage{amsfonts, amsmath, amsthm, amssymb} 
\usepackage{mathtools}
\usepackage{array}
\usepackage{soul}
\usepackage[colorlinks,bookmarks=false,citecolor=blue,linkcolor=red,urlcolor=blue]{hyperref}

\begin{document}
\title{Quantum frustrated Wigner chains}

\author{Rapha\"el Menu}
\affiliation{Theoretische  Physik,  Universit\"at  des  Saarlandes,  D-66123  Saarbr\"ucken,  Germany}
\author{Jorge Yago Malo}
\affiliation{Dipartimento di Fisica Enrico Fermi, Universita di Pisa and INFN, Largo B. Pontecorvo 3,
I-56127 Pisa, Italy.}
\author{Vladan Vuleti\'c}
\affiliation{Department of Physics, MIT-Harvard Center for Ultracold Atoms, and Research Laboratory of Electronics, Massachusetts Institute of Technology, Cambridge,
Massachusetts 02139, USA.}
\author{Maria Luisa Chiofalo}
\affiliation{Dipartimento di Fisica Enrico Fermi, Universita di Pisa and INFN, Largo B. Pontecorvo 3,
I-56127 Pisa, Italy.}
\author{Giovanna Morigi}
\affiliation{Theoretische  Physik,  Universit\"at  des  Saarlandes,  D-66123  Saarbr\"ucken,  Germany}

\date{\today}

\begin{abstract}
A Wigner chain in a periodic potential is a paradigmatic example of geometric frustration with long-range interactions. The dynamics emulates the Frenkel-Kontorova model with Coulomb interactions. In the continuum approximation, dislocations are sine-Gordon solitons with power-law decaying tails. We show that their action is mapped into a massive, long-range (1+1) Thirring model, where the solitons are charged fermionic excitations over an effective Dirac sea. We identify the corresponding mean field theory and show that the Coulomb interactions destabilize structures commensurate with the periodic substrate, suppressing their onset and giving rise to {\it interaction-induced} lubrication. Our study identifies the role of long-range interactions on determining nanofriction. Our predictions can be probed in state-of-the-art trapped ion experiments. 
\end{abstract}

\maketitle
{\it Introduction.}
Geometric frustration describes the impossibility for a system to simultaneously minimize all competing interactions due to its geometry \cite{Moessner:2003,Balents:2010}. %
The competition of ordering mechanisms often gives rise to phase transitions, such as the paradigmatic Aubry transition and the commensurate-incommensurate phase transition, separating a gapped ordered phase from a phase where the formation of defects becomes energetically favorable \cite{Pokrovkij_Talapov2,Braun_Kishvar}.
These phase transitions encompass several phenomena encountered in material science and condensed matter \cite{Bak_1982,Zwerger:2003,Tosatti:RMP} and are paradigmatic models for tribology \cite{Tosatti:RMP}. 

Several paradigms of frustration of condensed matter, where interactions are typically short-range, change substantially in the presence of long-range interactions. For two-body potentials $V(r)$ scaling with the distance $r$ as $V(r)\lesssim 1/r^{d}$, with $d$ the spatial dimensions, the energy becomes non-additive \cite{CAMPA200957} and domain walls are energetically unfavourable \cite{Botet:1982,Defenu:2018}. This gives rise to equilibrium and static properties which can be quite different from the short-range counterpart. In the transverse-field Ising model, for instance, the interplay of long-range interactions with quantum fluctuations confine excitations \cite{Liu:2019}. In the Bose-Hubbard model, global interactions give rise to exotic phases of ultracold matter \cite{Landig:2016,Habibian:2013,Sharma:2022}. Recent works predict that long-range, non-additive interactions can stabilize continuous symmetry breaking phases in the quantum XXZ model in one dimension \cite{Maghrebi:2017} and suppress the Berezinskii-Kosterlitz-Thouless (BKT) phase transition in the two-dimensional XY model, giving instead rise to an order-disorder phase transition as a function of the temperature \cite{Giachetti_2021,Defenu:2023}. Interestingly, the commensurate-incommensurate transition in one dimension and for short-range interactions is a BKT phase transition. In the context of tribology, this leads to the question of the role of long-range interactions in determining sliding. This question is important for material science and engineering and, at the same time, for our fundamental understanding of the interplay between frustration and long-range, non-additive forces. 

\begin{figure}
    \centering
    \includegraphics[width=0.9\columnwidth]{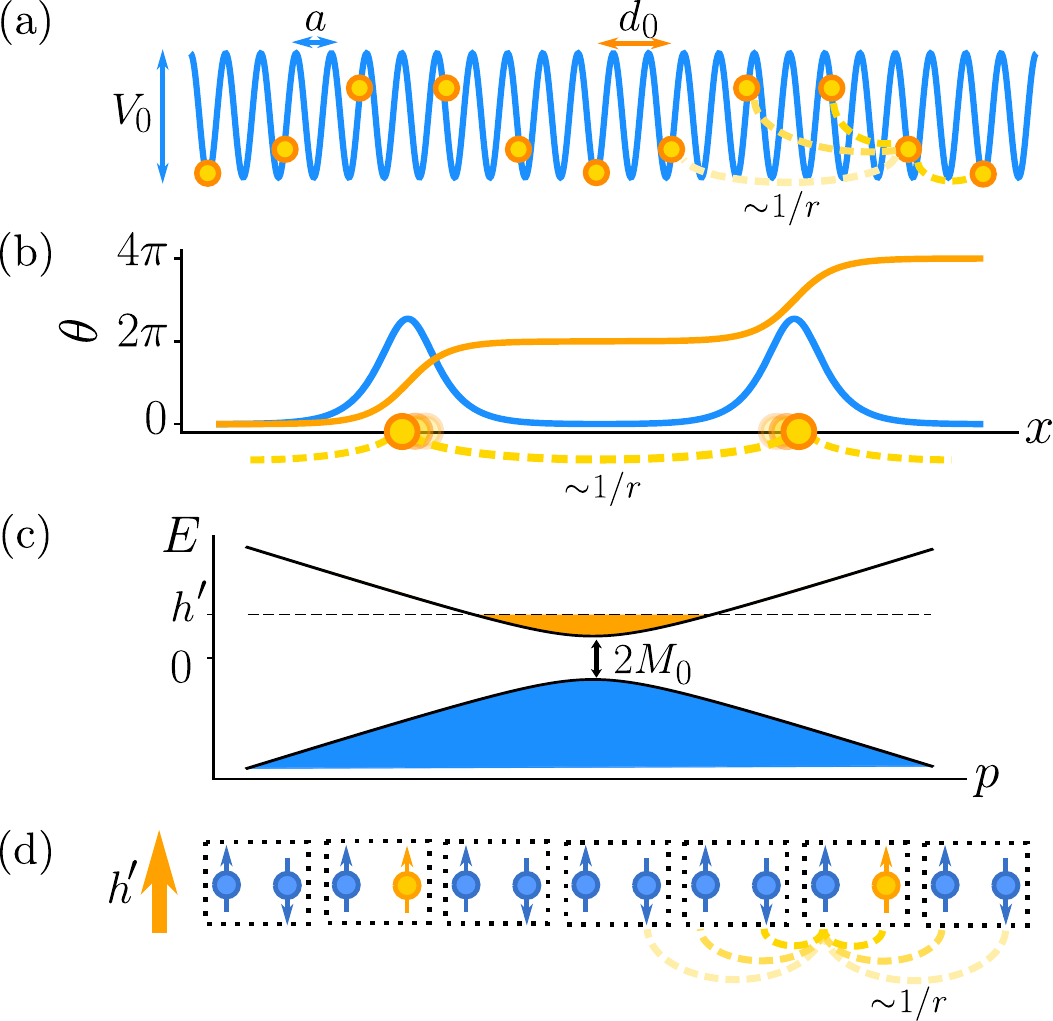}
    \caption{(a) Schematic representation of the Wigner crystal confined by an optical lattice. The substrate potential leads to dislocation of the ions from the crystal's equilibrium position. (b) The dislocations (solitons) are displayed as phase shift $\theta(x)$ along the chain axis $x$ (orange). The first-order derivative of $\theta(x)$ (blue) displays local maxima where the solitons are located. The solitons behave as interacting charges (yellow). (c) In the Thirring model the solitons are positive-energy excitation over a filled Dirac sea, where the gap is the soliton mass $M_0$. The phase is commensurate when the chemical potential $h'\propto \delta$ falls within the gap. For $h'>M_0$, the solitons proliferate and the phase is incommensurate. (d) Solitons are interacting spin defects (in orange) in the staggered ordering of a long-range interacting XXZ antiferromagnetic spin chain.}
    \label{fig:1}
\end{figure}

In this work, we unveil the tight connection among frustration, long-range interactions, and sliding in one dimension by analysing the ground-state properties of the Frenkel-Kontorova (FK) model with long-range interactions. 
The FK model describes an array of particles interacting with elastic forces and confined by a periodic potential, representing an underlying substrate \cite{Frenkel_Kontorova,Braun_Kishvar}. For nearest-neighbour interactions, its ground-state phase diagram includes the Aubry and the commensurate-incommensurate transition. The ground state is a non-analytic function of the mismatch between the substrate periodicity and the  array's characteristic length, taking the form of a devil's staircase of commensurate structures \cite{Bak_1982,Aubry:1983}. 
The transition to the incommensurate, sliding phase is discontinuous and occurs when the energy to create a dislocation (kink) vanishes, leading to kinks proliferation \cite{Bak_1982,Zwerger:2003,Pokrovskij_Talapov1}. In the continuum limit the kinks (see Fig. \ref{fig:1}(a)) are sine-Gordon solitons \cite{Merwe,Pokrovskij_Talapov1,Rubinstein:2003}. 

The FK model can be emulated by ultracold atomic gases \cite{Zwerger:2003,Dalmonte:2010,Kasper:2020}. Among the different platforms, chains of laser-cooled, self-organized arrays of ions in traps have been at the center of theoretical \cite{Garcia-Mata,Pruttivarasin:2011,Cormick:2013,Vanossi:2013,Cetina:2013}  and experimental studies \cite{Bylinskii:2015,Bylinskii:2016,Gangloff:2022,Kiethe:2017,Kiethe:2018}. The FK model is here realised by coupling the ions with an optical lattice, acting as substrate periodic potential \cite{Bylinskii:2015,Bylinskii:2016,Gangloff:2022}, or in a similar setting where two chains are sliding on top of each other \cite{Kiethe:2017,Kiethe:2018}. In these experiments the kinks statistics and dynamics could be measured \cite{Kiethe:2018,Gangloff:2022}, providing insightful information on the microscopic dynamics of the onset of friction. The experimental level of control opens the way towards studying quantum nanofriction, where tunnelling is expected to support lubrication \cite{TosattiPNAS,Mueser:2004,Mueser:2005,Bonetti,Timm:2021,Gangloff:2022,Chelpanova:2023}. The Coulomb interactions characteristic of trapped ions, however, is long-range and non-additive. These has important consequences: In the absence of a substrate, the decay of the correlations with the distance in one dimension is slower than a power law, and any finite chain is a one-dimensional Wigner crystal \cite{Schulz:1993}. In the presence of the substrate, the soliton's width (mass) exhibits an anomalous dependence on the chain size \cite{Landa:2020}, which contrasts with the finite, size-independent mass for power-law interactions of models with additive energy \cite{Dalmonte:2010}. Chains of charged particles, thus provide a platform for verifying theories and conjectures about the interplay of quantum fluctuations, long-range interactions, and frustration \cite{Maghrebi:2017,Giachetti_2021,Defenu:2023}. The Coulomb soliton's quantum dynamics and the nature of the corresponding commensurate-incommensurate phase transition are the objects of this study.

{\it Solitons in a frustrated Wigner chain.} We consider $N$ charged particles of mass $m$ and confined along a chain of length $L$. For periodic boundary conditions the Coulomb repulsion is minimized by an ordered structure of periodicity $d_0 = L/N$, where the ions are localized at the equilibrium positions $x_j^{(0)}=jd_0$ ($j=1,\ldots,N$). The Hamiltonian describing the harmonic vibrations about the equilibrium configuration reads \cite{Schulz:1993,Fishman:2008}
$$H_{\rm Wigner}=\sum_j\frac{p_j^2}{2m}+\frac{K}{2}\sum_j\sum_{r> 0}\frac{(x_{j+r}-x_j-rd_0)^2}{r^3}\,,$$
with $p_j$ and $x_j$ canonically conjugated variables and $K$ the stiffness. Frustration is introduced by a sinusoidal potential of depth $V_0$ and periodicity $a$, see Fig. \ref{fig:1}(a). When the interparticle distance is larger than the periodicity ($d_0\gg a$), the new Hamiltonian is $H_{\rm FK}=H_{\rm Wigner}-\sum_jV_0\cos(2\pi x_j/a)$ and the equations of motion describe a set of coupled pendula. Let $u_j$ denote the static particle displacement of the equilibrium positions $\overline{x}_j^{(0)}$ from the periodic ordering of a Wigner crystal $jd_0$, $u_j=\overline{x}_j^{(0)}-jd_0$. A finite value of the mismatch $\delta$ between Wigner and lattice periodicity breaks the discrete translational symmetry: $\delta=(d_0-n_0a)/a$ (with $n_0\in\mathbb N$ such that $0<\delta<1$). A dislocation is captured by the behavior of the phase $$\theta_j=\frac{2\pi}{a}(u_j + ja\delta)$$ as a function of $j$, see Fig.\ref{fig:1}(b).
In the continuum limit $\theta_j(t)\to \theta(x,t)$, where $x$ is the dimensionless position along the chain in units of the mean interparticle spacing $d_0$. The dynamics of the field $\theta(x,t)$ is governed by a modified sine-Gordon equation \cite{Pokrovsky_1983,Braun:1990,Landa:2020}
\begin{align}
    \dfrac{1}{v_s^2} \partial^2_t \theta &=  \partial^2_x\theta - M^2 \sin \theta \notag \\
    &+ \dfrac{1}{3}\partial_x\int_{1}^{N/2}{\dfrac{\partial_x\theta(x+u) + \partial_x\theta(x-u)}{u}\mathrm{d}u}\label{Eq1}\, ,
\end{align}
where the integral term accounts for the long-range Coulomb repulsion and is obtained by partial integration, after discarding edge effects (see Ref. \cite{Landa:2020} and Supplemental Material (SM) A \cite{SM}). 
The equation is parametrized by a velocity $v_s = \sqrt{{3K}/({2m})}$ and by a mass term $M = \sqrt{{8\pi^2V_0}/({3Ka^2})}$, the solutions shall satisfy the constraint given by the mismatch $\delta$.
In the short-range FK model the integral term vanishes and the solution is a sine-Gordon soliton \cite{Merwe,Pokrovskij_Talapov1,Rubinstein:2003}: $v_s$ is then the sound velocity and $M$ the soliton mass. The kink's length is proportional to $d_0/M$ and the validity of the continuum approximation requires $M\ll 1$. The action of the Sine-Gordon soliton can be mapped to the one of a solvable model of quantum field theory, the (1+1) massive Thirring model \cite{Thirring:1958,Coleman,Mandelstam:1975}, thereby connecting dislocations in a lattice with fermionic excitations over a Dirac vacuum. Here, the mismatch $\delta$ is an effective chemical potential and the commensurate-incommensurate transition a BKT \cite{Banuls_2020}. 

We now turn to the full Coulomb model. In the SM \cite{SM} we show that Eq.\ \eqref{Eq1} is the Euler-Lagrange equation of the action: 
\begin{align}
    &\mathcal{S}[\theta] = \dfrac{1}{\beta^2}\int{\mathrm{d} \tau \mathrm{d}x  \left[ \dfrac{1}{2}\left(\partial_\tau \theta\right)^2 - \dfrac{1}{2}\left(\partial_x\theta-2\pi\delta\right)^2 + M^2 \cos\theta \right.}\notag\\
    &- \left. \dfrac{1}{6}\int_{1}^{N/2}{\dfrac{\mathrm{d} u }{u}\left(\partial_x\theta(x)-2\pi\delta\right)\left(\partial_x\theta(x+u)+\partial_x\theta(x-u)-4\pi\delta\right)} \right]\,,\label{Eq2}
\end{align}
which differs from the action of the sine-Gordon model by an additional integral term, representing soliton-soliton Coulomb interactions \footnote{The action can be generalized to the case $d_0/a=n_0/m_0$, with $m_0>1$ and $n_0$ and $m_0$ prime to each other, see e.g.\ \cite{Pokrovskij_Talapov1}. Here, we restrict to the case $m_0=1$.}. This is strikingly different from the short-range case, where the solitons do not interact.  
The quantum field $\theta(x,\tau)$ and its conjugate ${\partial_\tau\theta}(x,\tau)$ satisfy the equal-time commutator $[\theta(x,\tau),\partial_\tau\theta(x',\tau) ]= i\beta^2\delta(x-x')$, where, $\tau=v_st$ is the rescaled time and $\beta^2 = (\frac{2\pi}{a})^2 \sqrt{\frac{2\hbar^2}{3mK}}$ is an effective Planck constant, corresponding to the ratio between kinetic and Coulomb characteristic energy. The dynamics is now fully determined by the soliton mass $M$, the effective Planck constant $\beta^2$, and the mismatch $\delta$.

{\it The long-range Thirring model.} In order to study the implications on the commensurate-incommensurate phase transition, we perform a mapping to a quantum lattice-gauge theory model. We use Mandelstam's definition of a soliton field  and consider the spinor $\psi^\dagger = (\psi_1^\dagger, \psi_2^\dagger)$ with components \cite{Mandelstam:1975}
    \begin{align}
        \psi_{j}(x) &= \dfrac{i^{(j-1)}}{\sqrt{2\pi}}\exp\left( - \dfrac{2\pi i}{\beta} \int_{-N/2}^x{{\partial_\tau\theta}(u)\mathrm{d} u } +(-1)^j \dfrac{i\beta}{2}\theta(x)\right),
    \end{align}
where now kink and anti-kink are fermion fields ($j=1,2$) with a different chirality. 
We linearize $\psi_{j}(x)$ for slowly-varying soliton fields ($|\partial_\tau{\theta}|, |\partial_x\theta| \ll 1$) and bring the action to the Hamiltonian form $H=\int {\rm d}x(\mathcal H(x)-h'\rho(x))$ where $\rho(x)=\psi^\dagger\psi$ is the fermions density \cite{Mandelstam:1975,SM} and $h'$ the chemical potential:
\begin{equation}
h'=\delta\left(\dfrac{2\pi}{\beta}\right)^2\left(1-\frac{2\ln 2}{3}+\frac{2}{3}\ln N\right)\,.
\end{equation}
As for the short-range model, the chemical potential is proportional to the mismatch. However, it now also includes the contribution of the Coulomb self-energy, which depends on $N$.
The Hamiltonian density $\mathcal H(x)$ is a long-range and massive (1+1) Thirring model:
\begin{eqnarray}
     &&\mathcal H(x) = -i c \overline{\psi} \gamma^1 \partial_x\psi + \dfrac{g}{4}(\overline{\psi}\gamma^\mu \psi)(\overline{\psi}\gamma_\mu \psi) + M_0 \overline{\psi}\psi \nonumber\\
     &&+\dfrac{(2\pi)^2}{6\beta^2}\int_{1}^{N/2}\dfrac{\mathrm{d}u}{u}\rho(x)\left(\rho(x + u) + \rho(x - u) \right)\label{Eq4},
\end{eqnarray}
with the adjoint spin $\overline{\psi}\equiv \psi^\dagger \gamma^0$  and $\mu=0,1$ such that $\gamma^0 = \sigma^z$, $\gamma^1 = i\sigma^y$. The dimensionless variables are the mass $M_0 = \frac{\pi M^2}{\beta^2}$ and the speed 
$c=\frac{2\pi}{\beta^2} + \frac{\beta^2}{8\pi}$. 
The fermions display Coulomb density-density interactions,  which scale as $N\ln N/\beta^2$, and contact interactions, scaling as $gN$ with $g = (\frac{2\pi}{\beta} )^2 - (\frac{\beta}{2})^2$. The mapping to the long-range Thirring Hamiltonian establishes an explicit link between frustration in a Wigner chain and a long-range quantum field theory. Now, the commensurate structure is gapped, with the filled Dirac sea of all negative-energy states, see Fig.\ \ref{fig:1}(c), and the chemical potential set below the gap, $h'<M_0$. The topological defects are charged fermionic excitations at positive energy, their energy cost is the rescaled soliton mass $M_0$. For $h'>M_0$ the energy for creating a soliton vanishes and the phase is incommensurate. The dependence on the chain size, $h'\sim \ln N$, indicates that the Coulomb interactions dominate, thus that it is energetically convenient to form a soliton. We thus expect no BKT transition as a function of $M_0$, in contrast with the nearest-neighbour case. 

{\it Thermodynamic limit.} The phases in the thermodynamic limit can be extracted by mapping the Thirring Hamiltonian $H$ into a lattice model of interacting fermions with a site-dependent chemical potential \cite{Susskind:1977,Banuls}. 
The procedure extends the one illustrated in Ref.\ \cite{Banuls_2020} to the long-range case and is detailed in SM~B \cite{SM}. The Thirring Hamiltonian is equivalent to a XXZ Hamiltonian of spin-$\frac{1}{2}$ particles in an external field: $\hat{H}_{\rm AFM}= \hat{H}_{\rm AFM}^{(0)}+\hat{H}_{\rm AFM}^{(c)}$, where $\hat{H}_{\rm AFM}^{(0)}$ is a nearest-neighbor XXZ model:
\begin{align}
    \hat{H}_{\rm AFM}^{(0)} &= - \dfrac{c}{2}\sum_n{(\hat{\sigma}^+_n \hat{\sigma}^-_{n+1} + \hat{\sigma}^-_n \hat{\sigma}^+_{n+1})}+ \dfrac{g}{2}\sum_n{\hat{\sigma}^z_n\hat{\sigma}\,,^z_{n+1}}\,,
\end{align}    
while the mismatch and Coulomb interactions are contained in the Hamiltonian term
\begin{align}
    \hat{H}_{\rm AFM}^{(c)} &= \sum_n{((-1)^nM_0-h')\hat{\sigma}^z_n}+ \dfrac{2\pi^2}{3\beta^2}\sum_n\sum_{r,\vert r\vert>1}{\dfrac{1}{\vert r \vert}\hat{\sigma}^z_n\hat{\sigma}^z_{n+r}}\, .
    \label{XXZ}
\end{align}
Hamiltonian $\hat{H}_{\rm AFM}$ describes the full extent of the quantum behaviour of a Wigner crystal in contact with a substrate. It encompasses the commensurate-incommensurate transition heralding the reorganization of the charged particles to fit the substrate's periodicity, with control field $h'$. In turn, the amplitude $M_0$ of the staggered field controls the Aubry transition signaling the pinning of the crystal to its substrate.  Now, for $h'=0$ the substrate imposes staggered order of the spins and the phase is an antiferromagnet. The homogeneous magnetic field $h'$(the mismatch) tends to flip spins in the up-oriented direction, therefore introducing defects (solitons) in the spin chain, as illustrated in the lower panel of Fig. \ref{fig:1}(d). 

The scaling of the terms in Eq. \eqref{XXZ} with $N$ is crucial. 
In fact, the non-additivity of the Coulomb interactions leads to the scaling $N\ln N$. We apply Kac's scaling, which restores the energy extensivity of long-range systems \cite{CAMPA200957}. This is equivalent to the prescription $K=K_0/\ln N$, which warrants that the Coulomb energy $\sum_{i,j}q^2/(d_0|i-j|)\sim K N\ln N=K_0 N$ becomes extensive in the thermodynamic limit \cite{Morigi:2004,Landa:2020}. As a consequence, $\beta^2=\beta_0^2\sqrt{\ln N}$ with $\beta_0^2=(\frac{2\pi}{a})^2 \sqrt{\frac{2\hbar^2}{3mK_0}}$ the rescaled Planck's constant after Kac's prescription. For large chains, $N\gg 1$, then the Hamiltonian terms scale as $\hat{H}_{\rm AFM}^{(0)} \sim {\rm O}(\beta_0^2)$ and $\hat{H}_{\rm AFM}^{(c)}\sim {\rm O}(1/\beta_0^2)$, see SM C \cite{SM}. Thus, for $\beta_0^2\gg 1$ the ground state is the one of the short range XXZ Hamiltonian with fixed values of $c$ and $g$. In this regime quantum fluctuations dominate and  the phase is incommensurate. In the opposite limit, $\beta_0^2\ll 1$, interactions dominate. In this limit the short-range model predicts a fractal structure, corresponding to a Devil's staircase of commensurate phases as a function of the mismatch. We now analyse this limit for the Coulomb case. 

\begin{figure}
    \centering
    \includegraphics[width=0.9\columnwidth]{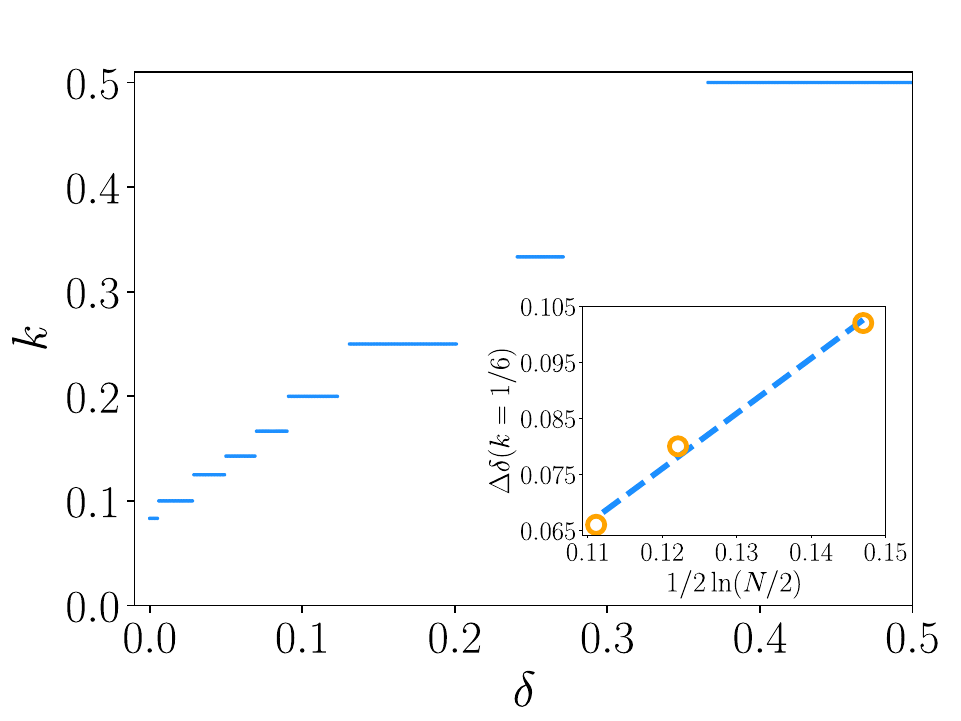}
    \caption{{The range of stability of some commensurate phases of Eq.\ \eqref{Eq:Bak} as a function of the mismatch $\delta$ forms a devil's staircase. The commensurate phases are here some fraction $k=1/n$ of up-oriented spins, with $n\in \mathbb{N}$.} Inset: the width of the plateau (here for the fraction $k=1/6$) decreases as a function of $\ln N$, showing that the steps of the staircase shrink in the thermodynamic limit. The fraction is computed using the method of Ref. \cite{Koziol:2023} for a chain of $N_s = 200$ spins and for $\tilde{M} = 0.05$. }
    \label{fig:3}
\end{figure}
{\it Devil's staircase.} Hamiltonian \eqref{XXZ} is diagonal in the basis of  $\sigma_z^j+1$, the eigenstates are strings $\{\sigma_j\}$ with $\sigma_j = 0,1$ with eigenvalues: 
\begin{align}
    E_{\rm mf}[\{\sigma_j\}] &= \dfrac{1}{2}\sum_{i,j}\frac{J_0}{|i-j|}{\sigma}_i {\sigma}_j -  \sum_i\tilde{B}_i{\sigma_i}\,,
    \label{Eq:Bak}
\end{align}
\noindent The commensurate configurations are here characterized by the density $k=1/n$ of up-oriented spins (magnetization) with $n\in\mathbb N$ and their stability are determined by the interplay of antiferromagnetic interactions, with amplitude $J_0=\gamma/(2\ln N)$, and a magnetic field $\tilde{B}_i=\gamma(\delta-\tilde{M}(-1)^i)$, with $\gamma=8\pi^2/(3\beta_0^2)$ a scaling factor and $\tilde{M}=V_0/(K_0 a^2)\propto M^2$. At fixed mismatch and varying potential depth, the field $\tilde B_i$ controls the Aubry transition from sliding to pinned phase \cite{Aubry:1983,Biham:1989}. Instead, at fixed potential depth $V_0$, $\tilde B_i$ controls the transition to an incommensurate structure. Figure \ref{fig:3} shows that the commensurate configurations form a devil's staircase as a function of the mismatch $\delta$. This is consistent with the prediction of analogous antiferromagnetic spin models \cite{Bak-PRL:1982,Bruisma:1983a}. The inset of Fig.\ \ref{fig:3} shows that the plateaus decrease as $1/\ln N$ suggesting that the staircase disappears in the thermodynamic limit. We confirm this behavior following the argument of Ref.\ \cite{Bak-PRL:1982}. Consider a commensurate configuration at $\delta=0$ with magnetization $k=r/n$ with $r,n\in\mathbb N$. The transition to an incommensurate structure occurs for the values of $\delta$ at which flipping a spin has zero energy cost, and occurs at the minimum $\delta_{\rm min}$ and maximum value $\delta_{\rm max}$ of the mismatch, while for  $\delta\in(\delta_{\rm min},\delta_{\rm max})$ the ground state is gapped and the commensurate phase stable. The interval's width $\Delta_\delta=\delta_{\rm max}-\delta_{\rm min}$ is the size of the plateau, it depends on $k=r/n$, and for 
$n$ even it takes the form 
\begin{equation}
   \Delta_\delta \approx\frac{1}{\ln N}\sum_{p=1}^{+\infty} \frac{1}{p^2n^2-1}\approx \frac{\zeta(2)}{n^2\ln N}\,.
\end{equation}
This demonstrates that the gap of the commensurate phases scale with $1/\ln N$ and the devil's staircase disappears in the thermodynamic limit.  This result is consistent with the result of \cite{Bruinsma:1983}, indicating that the fractal dimension of the ground state of Eq. \eqref{Eq:Bak} is unity. Hence, Coulomb interactions destabilize the commensurate structure, inducing sliding. Remarkably, defects are effective charges that interact suggesting that the gapless, sliding phase can exhibit long-range order.

{\it Experimental realizations.} Our study shows that the phase diagram is determined by the effective Planck's constant $\beta^2$. In ion chains $\beta^2$ can be tuned by changing the ratio $a/d_0$ and can take values ranging from $\sim 10^{-4}$, deep in the interaction-induced sliding phase, to $\sim 0.1$, where quantum lubrication is expected. Since the size of the commensurate phase vanishes with $1/\ln N$, a devil's staircase is measurable in any finite chain when the temperature $T$ is smaller than the gap. In Ref.\ \cite{Menu:2024} we report for a theoretical analysis based on the discrete charge distribution which confirms these results. Using the parameters of Ref. \cite{Bylinskii:2016}, for a chain of 100 $^{173}$Yb$^+$ ions  with interparticle distance $d_0 = 6 \mu\mathrm{m}$ and lattice periodicity $a=185\mathrm{nm}$, steps of the Devil's staircase with magnetization $k=1/n$ shall be visible for $T\lesssim 1 \,\mathrm{mK}/n$ \cite{Menu:2024} and features of quantum tunnelling of the solitons \cite{Timm:2021} can be observed in the spectrum.  One assumption of our study is the equidistance of the ions. This is realized in ring traps \cite{Haeffner} or in the central region of large ion chains in linear Paul traps \cite{Dubin:1997} (see \cite{Fogarty:2015} for the definition of the mismatch). We note, nevertheless, that a numerical study for 5 ions in a linear Paul trap \cite{Bonetti} is qualitatively consistent with our predictions. 

{\it Conclusions.} We have derived a theoretical framework that allows us to systematically account for geometric frustration, Coulomb interactions, and quantum fluctuations, exploring the long-range regime that separates the additive from the non-additive energy model and eluded previous theoretical investigation on frustration with long-range potentials \cite{Maghrebi:2017,Giachetti_2021,Defenu:2023}. Our model permits us to identify the relevant parameters and to show that long-range, non-additive interactions destabilize commensurate structures in one-dimension, giving rise to interaction-induced lubrication. Our predictions can be probed in systems with trapped ion chains. This framework paves the way for studying the nature of defects in one dimensional long-range interacting systems, finding relevant connections with lattice-gauge theoretical models and with the fractional quantum Hall effect  \cite{Rotondo_2016,Nussinov:2016,Petrova:2024}, as well as enabling their simulation with ion chains. 

{\it Acknowledgments}
The authors acknowledge discussions with O. Chelpanova, S. B. J\"ager, S. Kelly, J. Koziol, H. Landa, J. Marino, T. Mehlst\"aubler, and F. Schmidt-Kaler. 
This work was supported by the Deutsche Forschungsgemeinschaft (DFG, German Research Foundation), with the CRC-TRR 306 ``QuCoLiMa'', Project-ID No. 429529648 (B01),  by the German Ministry of Education and Research (BMBF) via the project NiQ ("Noise in Quantum Algorithm"), and in part by the National Science Foundation under Grants No. NSF PHY-1748958 and PHY-2309135. J.Y.M. was supported by the European Social Fund REACT EU through the Italian national program PON 2014-2020, DM MUR 1062/2021. M.L.C. acknowledges support from the National Centre on HPC, Big Data and Quantum Computing - SPOKE 10 (Quantum Computing) and received funding from the European Union Next-GenerationEU - National Recovery and Resilience Plan (NRRP) – MISSION 4 COMPONENT 2, INVESTMENT N. 1.4 – CUP N. I53C22000690001. M.L.C. acknowledge support from the project PRA\_2022\_2023\_98 "IMAGINATION" and acknowledges support from the MIT-UNIPI program.

\bibliography{biblio}

\begin{thebibliography}{67}%
\makeatletter
\providecommand \@ifxundefined [1]{%
 \@ifx{#1\undefined}
}%
\providecommand \@ifnum [1]{%
 \ifnum #1\expandafter \@firstoftwo
 \else \expandafter \@secondoftwo
 \fi
}%
\providecommand \@ifx [1]{%
 \ifx #1\expandafter \@firstoftwo
 \else \expandafter \@secondoftwo
 \fi
}%
\providecommand \natexlab [1]{#1}%
\providecommand \enquote  [1]{``#1''}%
\providecommand \bibnamefont  [1]{#1}%
\providecommand \bibfnamefont [1]{#1}%
\providecommand \citenamefont [1]{#1}%
\providecommand \href@noop [0]{\@secondoftwo}%
\providecommand \href [0]{\begingroup \@sanitize@url \@href}%
\providecommand \@href[1]{\@@startlink{#1}\@@href}%
\providecommand \@@href[1]{\endgroup#1\@@endlink}%
\providecommand \@sanitize@url [0]{\catcode `\\12\catcode `\$12\catcode
  `\&12\catcode `\#12\catcode `\^12\catcode `\_12\catcode `\%12\relax}%
\providecommand \@@startlink[1]{}%
\providecommand \@@endlink[0]{}%
\providecommand \url  [0]{\begingroup\@sanitize@url \@url }%
\providecommand \@url [1]{\endgroup\@href {#1}{\urlprefix }}%
\providecommand \urlprefix  [0]{URL }%
\providecommand \Eprint [0]{\href }%
\providecommand \doibase [0]{http://dx.doi.org/}%
\providecommand \selectlanguage [0]{\@gobble}%
\providecommand \bibinfo  [0]{\@secondoftwo}%
\providecommand \bibfield  [0]{\@secondoftwo}%
\providecommand \translation [1]{[#1]}%
\providecommand \BibitemOpen [0]{}%
\providecommand \bibitemStop [0]{}%
\providecommand \bibitemNoStop [0]{.\EOS\space}%
\providecommand \EOS [0]{\spacefactor3000\relax}%
\providecommand \BibitemShut  [1]{\csname bibitem#1\endcsname}%
\let\auto@bib@innerbib\@empty
\bibitem [{\citenamefont {Moessner}\ and\ \citenamefont
  {Ramirez}(2006)}]{Moessner:2003}%
  \BibitemOpen
  \bibfield  {author} {\bibinfo {author} {\bibfnamefont {R.}~\bibnamefont
  {Moessner}}\ and\ \bibinfo {author} {\bibfnamefont {A.~P.}\ \bibnamefont
  {Ramirez}},\ }\bibfield  {title} {\enquote {\bibinfo {title} {Geometrical
  frustration},}\ }\href {\doibase https://doi.org/10.1063/1.2186278}
  {\bibfield  {journal} {\bibinfo  {journal} {Phys. Today}\ }\textbf {\bibinfo
  {volume} {59}},\ \bibinfo {pages} {24} (\bibinfo {year} {2006})}\BibitemShut
  {NoStop}%
\bibitem [{\citenamefont {Balents}(2010)}]{Balents:2010}%
  \BibitemOpen
  \bibfield  {author} {\bibinfo {author} {\bibfnamefont {L.}~\bibnamefont
  {Balents}},\ }\bibfield  {title} {\enquote {\bibinfo {title} {Spin liquids in
  frustrated magnets},}\ }\href {\doibase 10.1038/nature08917} {\bibfield
  {journal} {\bibinfo  {journal} {Nature}\ }\textbf {\bibinfo {volume} {464}},\
  \bibinfo {pages} {199--208} (\bibinfo {year} {2010})}\BibitemShut {NoStop}%
\bibitem [{\citenamefont {Pokrovskij}\ and\ \citenamefont
  {Talapov}(1978)}]{Pokrovkij_Talapov2}%
  \BibitemOpen
  \bibfield  {author} {\bibinfo {author} {\bibfnamefont {V.L.}\ \bibnamefont
  {Pokrovskij}}\ and\ \bibinfo {author} {\bibfnamefont {A.L.}\ \bibnamefont
  {Talapov}},\ }\bibfield  {title} {\enquote {\bibinfo {title} {Phase
  transitions and vibrational spectra of almost commensurate structures},}\
  }\href@noop {} {\bibfield  {journal} {\bibinfo  {journal} {Zh. Eksp. Teor.
  Fiz.}\ }\textbf {\bibinfo {volume} {75}},\ \bibinfo {pages} {1151--1157}
  (\bibinfo {year} {1978})}\BibitemShut {NoStop}%
\bibitem [{\citenamefont {Braun}\ and\ \citenamefont
  {Kivshar}(2004)}]{Braun_Kishvar}%
  \BibitemOpen
  \bibfield  {author} {\bibinfo {author} {\bibfnamefont {O.M.}\ \bibnamefont
  {Braun}}\ and\ \bibinfo {author} {\bibfnamefont {Y.S.}\ \bibnamefont
  {Kivshar}},\ }\href@noop {} {\emph {\bibinfo {title} {The Frenkel-Kontorova
  Model: Concepts, Methods, and Applications}}}\ (\bibinfo  {publisher}
  {Springer},\ \bibinfo {address} {New York},\ \bibinfo {year}
  {2004})\BibitemShut {NoStop}%
\bibitem [{\citenamefont {Bak}(1982)}]{Bak_1982}%
  \BibitemOpen
  \bibfield  {author} {\bibinfo {author} {\bibfnamefont {P.}~\bibnamefont
  {Bak}},\ }\bibfield  {title} {\enquote {\bibinfo {title} {Commensurate
  phases, incommensurate phases and the devil's staircase},}\ }\href {\doibase
  10.1088/0034-4885/45/6/001} {\bibfield  {journal} {\bibinfo  {journal}
  {Reports on Progress in Physics}\ }\textbf {\bibinfo {volume} {45}},\
  \bibinfo {pages} {587--629} (\bibinfo {year} {1982})}\BibitemShut {NoStop}%
\bibitem [{\citenamefont {B\"uchler}\ \emph {et~al.}(2003)\citenamefont
  {B\"uchler}, \citenamefont {Blatter},\ and\ \citenamefont
  {Zwerger}}]{Zwerger:2003}%
  \BibitemOpen
  \bibfield  {author} {\bibinfo {author} {\bibfnamefont {H.~P.}\ \bibnamefont
  {B\"uchler}}, \bibinfo {author} {\bibfnamefont {G.}~\bibnamefont {Blatter}},
  \ and\ \bibinfo {author} {\bibfnamefont {W.}~\bibnamefont {Zwerger}},\
  }\bibfield  {title} {\enquote {\bibinfo {title} {Commensurate-incommensurate
  transition of cold atoms in an optical lattice},}\ }\href {\doibase
  10.1103/PhysRevLett.90.130401} {\bibfield  {journal} {\bibinfo  {journal}
  {Phys. Rev. Lett.}\ }\textbf {\bibinfo {volume} {90}},\ \bibinfo {pages}
  {130401} (\bibinfo {year} {2003})}\BibitemShut {NoStop}%
\bibitem [{\citenamefont {Wang}\ \emph {et~al.}(2024)\citenamefont {Wang},
  \citenamefont {Khosravi}, \citenamefont {Vanossi},\ and\ \citenamefont
  {Tosatti}}]{Tosatti:RMP}%
  \BibitemOpen
  \bibfield  {author} {\bibinfo {author} {\bibfnamefont {J.}~\bibnamefont
  {Wang}}, \bibinfo {author} {\bibfnamefont {A.}~\bibnamefont {Khosravi}},
  \bibinfo {author} {\bibfnamefont {A.}~\bibnamefont {Vanossi}}, \ and\
  \bibinfo {author} {\bibfnamefont {E.}~\bibnamefont {Tosatti}},\ }\bibfield
  {title} {\enquote {\bibinfo {title} {Colloquium: Sliding and pinning in
  structurally lubric 2d material interfaces},}\ }\href {\doibase
  10.1103/RevModPhys.96.011002} {\bibfield  {journal} {\bibinfo  {journal}
  {Rev. Mod. Phys.}\ }\textbf {\bibinfo {volume} {96}},\ \bibinfo {pages}
  {011002} (\bibinfo {year} {2024})}\BibitemShut {NoStop}%
\bibitem [{\citenamefont {Campa}\ \emph {et~al.}(2009)\citenamefont {Campa},
  \citenamefont {Dauxois},\ and\ \citenamefont {Ruffo}}]{CAMPA200957}%
  \BibitemOpen
  \bibfield  {author} {\bibinfo {author} {\bibfnamefont {A.}~\bibnamefont
  {Campa}}, \bibinfo {author} {\bibfnamefont {T.}~\bibnamefont {Dauxois}}, \
  and\ \bibinfo {author} {\bibfnamefont {S.}~\bibnamefont {Ruffo}},\ }\bibfield
   {title} {\enquote {\bibinfo {title} {Statistical mechanics and dynamics of
  solvable models with long-range interactions},}\ }\href {\doibase
  https://doi.org/10.1016/j.physrep.2009.07.001} {\bibfield  {journal}
  {\bibinfo  {journal} {Physics Reports}\ }\textbf {\bibinfo {volume} {480}},\
  \bibinfo {pages} {57--159} (\bibinfo {year} {2009})}\BibitemShut {NoStop}%
\bibitem [{\citenamefont {Botet}\ \emph {et~al.}(1982)\citenamefont {Botet},
  \citenamefont {Jullien},\ and\ \citenamefont {Pfeuty}}]{Botet:1982}%
  \BibitemOpen
  \bibfield  {author} {\bibinfo {author} {\bibfnamefont {R.}~\bibnamefont
  {Botet}}, \bibinfo {author} {\bibfnamefont {R.}~\bibnamefont {Jullien}}, \
  and\ \bibinfo {author} {\bibfnamefont {P.}~\bibnamefont {Pfeuty}},\
  }\bibfield  {title} {\enquote {\bibinfo {title} {Size scaling for infinitely
  coordinated systems},}\ }\href {\doibase 10.1103/PhysRevLett.49.478}
  {\bibfield  {journal} {\bibinfo  {journal} {Phys. Rev. Lett.}\ }\textbf
  {\bibinfo {volume} {49}},\ \bibinfo {pages} {478--481} (\bibinfo {year}
  {1982})}\BibitemShut {NoStop}%
\bibitem [{\citenamefont {Defenu}\ \emph {et~al.}(2018)\citenamefont {Defenu},
  \citenamefont {Enss}, \citenamefont {Kastner},\ and\ \citenamefont
  {Morigi}}]{Defenu:2018}%
  \BibitemOpen
  \bibfield  {author} {\bibinfo {author} {\bibfnamefont {N.}~\bibnamefont
  {Defenu}}, \bibinfo {author} {\bibfnamefont {T.}~\bibnamefont {Enss}},
  \bibinfo {author} {\bibfnamefont {M.}~\bibnamefont {Kastner}}, \ and\
  \bibinfo {author} {\bibfnamefont {G.}~\bibnamefont {Morigi}},\ }\bibfield
  {title} {\enquote {\bibinfo {title} {Dynamical critical scaling of long-range
  interacting quantum magnets},}\ }\href {\doibase
  10.1103/PhysRevLett.121.240403} {\bibfield  {journal} {\bibinfo  {journal}
  {Phys. Rev. Lett.}\ }\textbf {\bibinfo {volume} {121}},\ \bibinfo {pages}
  {240403} (\bibinfo {year} {2018})}\BibitemShut {NoStop}%
\bibitem [{\citenamefont {Liu}\ \emph {et~al.}(2019)\citenamefont {Liu},
  \citenamefont {Lundgren}, \citenamefont {Titum}, \citenamefont {Pagano},
  \citenamefont {Zhang}, \citenamefont {Monroe},\ and\ \citenamefont
  {Gorshkov}}]{Liu:2019}%
  \BibitemOpen
  \bibfield  {author} {\bibinfo {author} {\bibfnamefont {F.}~\bibnamefont
  {Liu}}, \bibinfo {author} {\bibfnamefont {R.}~\bibnamefont {Lundgren}},
  \bibinfo {author} {\bibfnamefont {P.}~\bibnamefont {Titum}}, \bibinfo
  {author} {\bibfnamefont {G.}~\bibnamefont {Pagano}}, \bibinfo {author}
  {\bibfnamefont {J.}~\bibnamefont {Zhang}}, \bibinfo {author} {\bibfnamefont
  {C.}~\bibnamefont {Monroe}}, \ and\ \bibinfo {author} {\bibfnamefont {A.~V.}\
  \bibnamefont {Gorshkov}},\ }\bibfield  {title} {\enquote {\bibinfo {title}
  {Confined quasiparticle dynamics in long-range interacting quantum spin
  chains},}\ }\href {\doibase 10.1103/PhysRevLett.122.150601} {\bibfield
  {journal} {\bibinfo  {journal} {Phys. Rev. Lett.}\ }\textbf {\bibinfo
  {volume} {122}},\ \bibinfo {pages} {150601} (\bibinfo {year}
  {2019})}\BibitemShut {NoStop}%
\bibitem [{\citenamefont {Landig}\ \emph {et~al.}(2016)\citenamefont {Landig},
  \citenamefont {Hruby}, \citenamefont {Dogra}, \citenamefont {Landini},
  \citenamefont {Mottl}, \citenamefont {Donner},\ and\ \citenamefont
  {Esslinger}}]{Landig:2016}%
  \BibitemOpen
  \bibfield  {author} {\bibinfo {author} {\bibfnamefont {R.}~\bibnamefont
  {Landig}}, \bibinfo {author} {\bibfnamefont {L.}~\bibnamefont {Hruby}},
  \bibinfo {author} {\bibfnamefont {N.}~\bibnamefont {Dogra}}, \bibinfo
  {author} {\bibfnamefont {M.}~\bibnamefont {Landini}}, \bibinfo {author}
  {\bibfnamefont {R.}~\bibnamefont {Mottl}}, \bibinfo {author} {\bibfnamefont
  {T.}~\bibnamefont {Donner}}, \ and\ \bibinfo {author} {\bibfnamefont
  {T.}~\bibnamefont {Esslinger}},\ }\bibfield  {title} {\enquote {\bibinfo
  {title} {Quantum phases from competing short- and long-range interactions in
  an optical lattice},}\ }\href {\doibase 10.1038/nature17409} {\bibfield
  {journal} {\bibinfo  {journal} {Nature}\ }\textbf {\bibinfo {volume} {532}},\
  \bibinfo {pages} {476--479} (\bibinfo {year} {2016})}\BibitemShut {NoStop}%
\bibitem [{\citenamefont {Habibian}\ \emph {et~al.}(2013)\citenamefont
  {Habibian}, \citenamefont {Winter}, \citenamefont {Paganelli}, \citenamefont
  {Rieger},\ and\ \citenamefont {Morigi}}]{Habibian:2013}%
  \BibitemOpen
  \bibfield  {author} {\bibinfo {author} {\bibfnamefont {H.}~\bibnamefont
  {Habibian}}, \bibinfo {author} {\bibfnamefont {A.}~\bibnamefont {Winter}},
  \bibinfo {author} {\bibfnamefont {S.}~\bibnamefont {Paganelli}}, \bibinfo
  {author} {\bibfnamefont {H.}~\bibnamefont {Rieger}}, \ and\ \bibinfo {author}
  {\bibfnamefont {G.}~\bibnamefont {Morigi}},\ }\bibfield  {title} {\enquote
  {\bibinfo {title} {Bose-glass phases of ultracold atoms due to cavity
  backaction},}\ }\href {\doibase 10.1103/PhysRevLett.110.075304} {\bibfield
  {journal} {\bibinfo  {journal} {Phys. Rev. Lett.}\ }\textbf {\bibinfo
  {volume} {110}},\ \bibinfo {pages} {075304} (\bibinfo {year}
  {2013})}\BibitemShut {NoStop}%
\bibitem [{\citenamefont {Sharma}\ \emph {et~al.}(2022)\citenamefont {Sharma},
  \citenamefont {J\"ager}, \citenamefont {Kraus}, \citenamefont {Roscilde},\
  and\ \citenamefont {Morigi}}]{Sharma:2022}%
  \BibitemOpen
  \bibfield  {author} {\bibinfo {author} {\bibfnamefont {S.}~\bibnamefont
  {Sharma}}, \bibinfo {author} {\bibfnamefont {S.~B.}\ \bibnamefont {J\"ager}},
  \bibinfo {author} {\bibfnamefont {R.}~\bibnamefont {Kraus}}, \bibinfo
  {author} {\bibfnamefont {T.}~\bibnamefont {Roscilde}}, \ and\ \bibinfo
  {author} {\bibfnamefont {G.}~\bibnamefont {Morigi}},\ }\bibfield  {title}
  {\enquote {\bibinfo {title} {Quantum critical behavior of entanglement in
  lattice bosons with cavity-mediated long-range interactions},}\ }\href
  {\doibase 10.1103/PhysRevLett.129.143001} {\bibfield  {journal} {\bibinfo
  {journal} {Phys. Rev. Lett.}\ }\textbf {\bibinfo {volume} {129}},\ \bibinfo
  {pages} {143001} (\bibinfo {year} {2022})}\BibitemShut {NoStop}%
\bibitem [{\citenamefont {Maghrebi}\ \emph {et~al.}(2017)\citenamefont
  {Maghrebi}, \citenamefont {Gong},\ and\ \citenamefont
  {Gorshkov}}]{Maghrebi:2017}%
  \BibitemOpen
  \bibfield  {author} {\bibinfo {author} {\bibfnamefont {M.~F.}\ \bibnamefont
  {Maghrebi}}, \bibinfo {author} {\bibfnamefont {Z.-X.}\ \bibnamefont {Gong}},
  \ and\ \bibinfo {author} {\bibfnamefont {A.~V.}\ \bibnamefont {Gorshkov}},\
  }\bibfield  {title} {\enquote {\bibinfo {title} {Continuous symmetry breaking
  in 1d long-range interacting quantum systems},}\ }\href {\doibase
  10.1103/PhysRevLett.119.023001} {\bibfield  {journal} {\bibinfo  {journal}
  {Phys. Rev. Lett.}\ }\textbf {\bibinfo {volume} {119}},\ \bibinfo {pages}
  {023001} (\bibinfo {year} {2017})}\BibitemShut {NoStop}%
\bibitem [{\citenamefont {Giachetti}\ \emph {et~al.}(2021)\citenamefont
  {Giachetti}, \citenamefont {Defenu}, \citenamefont {Ruffo},\ and\
  \citenamefont {Trombettoni}}]{Giachetti_2021}%
  \BibitemOpen
  \bibfield  {author} {\bibinfo {author} {\bibfnamefont {G.}~\bibnamefont
  {Giachetti}}, \bibinfo {author} {\bibfnamefont {N.}~\bibnamefont {Defenu}},
  \bibinfo {author} {\bibfnamefont {S.}~\bibnamefont {Ruffo}}, \ and\ \bibinfo
  {author} {\bibfnamefont {A.}~\bibnamefont {Trombettoni}},\ }\bibfield
  {title} {\enquote {\bibinfo {title} {Berezinskii-kosterlitz-thouless phase
  transitions with long-range couplings},}\ }\href {\doibase
  10.1103/PhysRevLett.127.156801} {\bibfield  {journal} {\bibinfo  {journal}
  {Phys. Rev. Lett.}\ }\textbf {\bibinfo {volume} {127}},\ \bibinfo {pages}
  {156801} (\bibinfo {year} {2021})}\BibitemShut {NoStop}%
\bibitem [{\citenamefont {Defenu}\ \emph {et~al.}(2023)\citenamefont {Defenu},
  \citenamefont {Donner}, \citenamefont {Macr\`{\i}}, \citenamefont {Pagano},
  \citenamefont {Ruffo},\ and\ \citenamefont {Trombettoni}}]{Defenu:2023}%
  \BibitemOpen
  \bibfield  {author} {\bibinfo {author} {\bibfnamefont {N.}~\bibnamefont
  {Defenu}}, \bibinfo {author} {\bibfnamefont {T.}~\bibnamefont {Donner}},
  \bibinfo {author} {\bibfnamefont {T.}~\bibnamefont {Macr\`{\i}}}, \bibinfo
  {author} {\bibfnamefont {G.}~\bibnamefont {Pagano}}, \bibinfo {author}
  {\bibfnamefont {S.}~\bibnamefont {Ruffo}}, \ and\ \bibinfo {author}
  {\bibfnamefont {A.}~\bibnamefont {Trombettoni}},\ }\bibfield  {title}
  {\enquote {\bibinfo {title} {Long-range interacting quantum systems},}\
  }\href {\doibase 10.1103/RevModPhys.95.035002} {\bibfield  {journal}
  {\bibinfo  {journal} {Rev. Mod. Phys.}\ }\textbf {\bibinfo {volume} {95}},\
  \bibinfo {pages} {035002} (\bibinfo {year} {2023})}\BibitemShut {NoStop}%
\bibitem [{\citenamefont {Frenkel}\ and\ \citenamefont
  {Kontorova}(1938)}]{Frenkel_Kontorova}%
  \BibitemOpen
  \bibfield  {author} {\bibinfo {author} {\bibfnamefont {Y.I.}\ \bibnamefont
  {Frenkel}}\ and\ \bibinfo {author} {\bibfnamefont {T.A.}\ \bibnamefont
  {Kontorova}},\ }\bibfield  {title} {\enquote {\bibinfo {title} {The model of
  dislocation in solid body},}\ }\href@noop {} {\bibfield  {journal} {\bibinfo
  {journal} {Zh. Eksp. Teor. Fiz.}\ }\textbf {\bibinfo {volume} {8}},\ \bibinfo
  {pages} {1340} (\bibinfo {year} {1938})}\BibitemShut {NoStop}%
\bibitem [{\citenamefont {Aubry}\ and\ \citenamefont {{Le
  Daeron}}(1983)}]{Aubry:1983}%
  \BibitemOpen
  \bibfield  {author} {\bibinfo {author} {\bibfnamefont {S.}~\bibnamefont
  {Aubry}}\ and\ \bibinfo {author} {\bibfnamefont {P.Y.}\ \bibnamefont {{Le
  Daeron}}},\ }\bibfield  {title} {\enquote {\bibinfo {title} {The discrete
  frenkel-kontorova model and its extensions: I. exact results for the
  ground-states},}\ }\href {\doibase
  https://doi.org/10.1016/0167-2789(83)90233-6} {\bibfield  {journal} {\bibinfo
   {journal} {Physica D: Nonlinear Phenomena}\ }\textbf {\bibinfo {volume}
  {8}},\ \bibinfo {pages} {381--422} (\bibinfo {year} {1983})}\BibitemShut
  {NoStop}%
\bibitem [{\citenamefont {Pokrovskij}\ and\ \citenamefont
  {Talapov}(1984)}]{Pokrovskij_Talapov1}%
  \BibitemOpen
  \bibfield  {author} {\bibinfo {author} {\bibfnamefont {V.L.}\ \bibnamefont
  {Pokrovskij}}\ and\ \bibinfo {author} {\bibfnamefont {A.L.}\ \bibnamefont
  {Talapov}},\ }\href@noop {} {\emph {\bibinfo {title} {Theory of
  incommensurate crystals}}}\ (\bibinfo  {publisher} {Harwood Academic
  Publishers},\ \bibinfo {address} {New York},\ \bibinfo {year}
  {1984})\BibitemShut {NoStop}%
\bibitem [{\citenamefont {Frank}\ \emph {et~al.}(1949)\citenamefont {Frank},
  \citenamefont {van~der Merwe},\ and\ \citenamefont {Mott}}]{Merwe}%
  \BibitemOpen
  \bibfield  {author} {\bibinfo {author} {\bibfnamefont {F.~C.}\ \bibnamefont
  {Frank}}, \bibinfo {author} {\bibfnamefont {J.~H.}\ \bibnamefont {van~der
  Merwe}}, \ and\ \bibinfo {author} {\bibfnamefont {Nevill~Francis}\
  \bibnamefont {Mott}},\ }\bibfield  {title} {\enquote {\bibinfo {title}
  {One-dimensional dislocations. i. static theory},}\ }\href {\doibase
  10.1098/rspa.1949.0095} {\bibfield  {journal} {\bibinfo  {journal}
  {Proceedings of the Royal Society of London. Series A. Mathematical and
  Physical Sciences}\ }\textbf {\bibinfo {volume} {198}},\ \bibinfo {pages}
  {205--216} (\bibinfo {year} {1949})}\BibitemShut {NoStop}%
\bibitem [{\citenamefont {Rubinstein}(2003)}]{Rubinstein:2003}%
  \BibitemOpen
  \bibfield  {author} {\bibinfo {author} {\bibfnamefont {J.}~\bibnamefont
  {Rubinstein}},\ }\bibfield  {title} {\enquote {\bibinfo {title}
  {{Sine--Gordon Equation}},}\ }\href {\doibase 10.1063/1.1665057} {\bibfield
  {journal} {\bibinfo  {journal} {Journal of Mathematical Physics}\ }\textbf
  {\bibinfo {volume} {11}},\ \bibinfo {pages} {258--266} (\bibinfo {year}
  {2003})}\BibitemShut {NoStop}%
\bibitem [{\citenamefont {Dalmonte}\ \emph {et~al.}(2010)\citenamefont
  {Dalmonte}, \citenamefont {Pupillo},\ and\ \citenamefont
  {Zoller}}]{Dalmonte:2010}%
  \BibitemOpen
  \bibfield  {author} {\bibinfo {author} {\bibfnamefont {M.}~\bibnamefont
  {Dalmonte}}, \bibinfo {author} {\bibfnamefont {G.}~\bibnamefont {Pupillo}}, \
  and\ \bibinfo {author} {\bibfnamefont {P.}~\bibnamefont {Zoller}},\
  }\bibfield  {title} {\enquote {\bibinfo {title} {One-dimensional quantum
  liquids with power-law interactions: The luttinger staircase},}\ }\href
  {\doibase 10.1103/PhysRevLett.105.140401} {\bibfield  {journal} {\bibinfo
  {journal} {Phys. Rev. Lett.}\ }\textbf {\bibinfo {volume} {105}},\ \bibinfo
  {pages} {140401} (\bibinfo {year} {2010})}\BibitemShut {NoStop}%
\bibitem [{\citenamefont {Kasper}\ \emph {et~al.}(2020)\citenamefont {Kasper},
  \citenamefont {Marino}, \citenamefont {Ji}, \citenamefont {Gritsev},
  \citenamefont {Schmiedmayer},\ and\ \citenamefont {Demler}}]{Kasper:2020}%
  \BibitemOpen
  \bibfield  {author} {\bibinfo {author} {\bibfnamefont {V.}~\bibnamefont
  {Kasper}}, \bibinfo {author} {\bibfnamefont {J.}~\bibnamefont {Marino}},
  \bibinfo {author} {\bibfnamefont {S.}~\bibnamefont {Ji}}, \bibinfo {author}
  {\bibfnamefont {V.}~\bibnamefont {Gritsev}}, \bibinfo {author} {\bibfnamefont
  {J.}~\bibnamefont {Schmiedmayer}}, \ and\ \bibinfo {author} {\bibfnamefont
  {E.}~\bibnamefont {Demler}},\ }\bibfield  {title} {\enquote {\bibinfo {title}
  {Simulating a quantum commensurate-incommensurate phase transition using two
  raman-coupled one-dimensional condensates},}\ }\href {\doibase
  10.1103/PhysRevB.101.224102} {\bibfield  {journal} {\bibinfo  {journal}
  {Phys. Rev. B}\ }\textbf {\bibinfo {volume} {101}},\ \bibinfo {pages}
  {224102} (\bibinfo {year} {2020})}\BibitemShut {NoStop}%
\bibitem [{\citenamefont {Garc\'{\i}a-Mata}\ \emph {et~al.}(2007)\citenamefont
  {Garc\'{\i}a-Mata}, \citenamefont {Zhirov},\ and\ \citenamefont
  {Shepelyansky}}]{Garcia-Mata}%
  \BibitemOpen
  \bibfield  {author} {\bibinfo {author} {\bibfnamefont {I.}~\bibnamefont
  {Garc\'{\i}a-Mata}}, \bibinfo {author} {\bibfnamefont {O.~V.}\ \bibnamefont
  {Zhirov}}, \ and\ \bibinfo {author} {\bibfnamefont {D.~L.}\ \bibnamefont
  {Shepelyansky}},\ }\bibfield  {title} {\enquote {\bibinfo {title}
  {Frenkel-kontorova model with cold trapped ions},}\ }\href {\doibase
  10.1140/epjd/e2006-00220-2} {\bibfield  {journal} {\bibinfo  {journal} {Eur.
  Phys. J. D}\ }\textbf {\bibinfo {volume} {41}},\ \bibinfo {pages} {325--330}
  (\bibinfo {year} {2007})}\BibitemShut {NoStop}%
\bibitem [{\citenamefont {Pruttivarasin}\ \emph {et~al.}(2011)\citenamefont
  {Pruttivarasin}, \citenamefont {Ramm}, \citenamefont {Talukdar},
  \citenamefont {Kreuter},\ and\ \citenamefont
  {Häffner}}]{Pruttivarasin:2011}%
  \BibitemOpen
  \bibfield  {author} {\bibinfo {author} {\bibfnamefont {T.}~\bibnamefont
  {Pruttivarasin}}, \bibinfo {author} {\bibfnamefont {M.}~\bibnamefont {Ramm}},
  \bibinfo {author} {\bibfnamefont {I.}~\bibnamefont {Talukdar}}, \bibinfo
  {author} {\bibfnamefont {A.}~\bibnamefont {Kreuter}}, \ and\ \bibinfo
  {author} {\bibfnamefont {H.}~\bibnamefont {Häffner}},\ }\bibfield  {title}
  {\enquote {\bibinfo {title} {Trapped ions in optical lattices for probing
  oscillator chain models},}\ }\href {\doibase 10.1088/1367-2630/13/7/075012}
  {\bibfield  {journal} {\bibinfo  {journal} {New Journal of Physics}\ }\textbf
  {\bibinfo {volume} {13}},\ \bibinfo {pages} {075012} (\bibinfo {year}
  {2011})}\BibitemShut {NoStop}%
\bibitem [{\citenamefont {Cormick}\ and\ \citenamefont
  {Morigi}(2013)}]{Cormick:2013}%
  \BibitemOpen
  \bibfield  {author} {\bibinfo {author} {\bibfnamefont {C.}~\bibnamefont
  {Cormick}}\ and\ \bibinfo {author} {\bibfnamefont {G.}~\bibnamefont
  {Morigi}},\ }\bibfield  {title} {\enquote {\bibinfo {title} {Ion chains in
  high-finesse cavities},}\ }\href {\doibase 10.1103/PhysRevA.87.013829}
  {\bibfield  {journal} {\bibinfo  {journal} {Phys. Rev. A}\ }\textbf {\bibinfo
  {volume} {87}},\ \bibinfo {pages} {013829} (\bibinfo {year}
  {2013})}\BibitemShut {NoStop}%
\bibitem [{\citenamefont {Vanossi}\ \emph {et~al.}(2013)\citenamefont
  {Vanossi}, \citenamefont {Manini}, \citenamefont {Urbakh}, \citenamefont
  {Zapperi},\ and\ \citenamefont {Tosatti}}]{Vanossi:2013}%
  \BibitemOpen
  \bibfield  {author} {\bibinfo {author} {\bibfnamefont {A.}~\bibnamefont
  {Vanossi}}, \bibinfo {author} {\bibfnamefont {N.}~\bibnamefont {Manini}},
  \bibinfo {author} {\bibfnamefont {M.}~\bibnamefont {Urbakh}}, \bibinfo
  {author} {\bibfnamefont {S.}~\bibnamefont {Zapperi}}, \ and\ \bibinfo
  {author} {\bibfnamefont {E.}~\bibnamefont {Tosatti}},\ }\bibfield  {title}
  {\enquote {\bibinfo {title} {Colloquium: Modeling friction: From nanoscale to
  mesoscale},}\ }\href {\doibase 10.1103/RevModPhys.85.529} {\bibfield
  {journal} {\bibinfo  {journal} {Rev. Mod. Phys.}\ }\textbf {\bibinfo {volume}
  {85}},\ \bibinfo {pages} {529--552} (\bibinfo {year} {2013})}\BibitemShut
  {NoStop}%
\bibitem [{\citenamefont {Cetina}\ \emph {et~al.}(2013)\citenamefont {Cetina},
  \citenamefont {Bylinskii}, \citenamefont {Karpa}, \citenamefont {Gangloff},
  \citenamefont {Beck}, \citenamefont {Ge}, \citenamefont {Scholz},
  \citenamefont {Grier}, \citenamefont {Chuang},\ and\ \citenamefont
  {Vuleti\'{c}}}]{Cetina:2013}%
  \BibitemOpen
  \bibfield  {author} {\bibinfo {author} {\bibfnamefont {M.}~\bibnamefont
  {Cetina}}, \bibinfo {author} {\bibfnamefont {A.}~\bibnamefont {Bylinskii}},
  \bibinfo {author} {\bibfnamefont {L.}~\bibnamefont {Karpa}}, \bibinfo
  {author} {\bibfnamefont {D.}~\bibnamefont {Gangloff}}, \bibinfo {author}
  {\bibfnamefont {K.~M.}\ \bibnamefont {Beck}}, \bibinfo {author}
  {\bibfnamefont {Y.}~\bibnamefont {Ge}}, \bibinfo {author} {\bibfnamefont
  {M.}~\bibnamefont {Scholz}}, \bibinfo {author} {\bibfnamefont {A.~T.}\
  \bibnamefont {Grier}}, \bibinfo {author} {\bibfnamefont {I.}~\bibnamefont
  {Chuang}}, \ and\ \bibinfo {author} {\bibfnamefont {V.}~\bibnamefont
  {Vuleti\'{c}}},\ }\bibfield  {title} {\enquote {\bibinfo {title}
  {One-dimensional array of ion chains coupled to an optical cavity},}\ }\href
  {\doibase 10.1088/1367-2630/15/5/053001} {\bibfield  {journal} {\bibinfo
  {journal} {New Journal of Physics}\ }\textbf {\bibinfo {volume} {15}},\
  \bibinfo {pages} {053001} (\bibinfo {year} {2013})}\BibitemShut {NoStop}%
\bibitem [{\citenamefont {Bylinskii}\ \emph {et~al.}(2015)\citenamefont
  {Bylinskii}, \citenamefont {Gangloff},\ and\ \citenamefont
  {Vuleti{\'c}}}]{Bylinskii:2015}%
  \BibitemOpen
  \bibfield  {author} {\bibinfo {author} {\bibfnamefont {A.}~\bibnamefont
  {Bylinskii}}, \bibinfo {author} {\bibfnamefont {D.}~\bibnamefont {Gangloff}},
  \ and\ \bibinfo {author} {\bibfnamefont {V.}~\bibnamefont {Vuleti{\'c}}},\
  }\bibfield  {title} {\enquote {\bibinfo {title} {Tuning friction atom-by-atom
  in an ion-crystal simulator},}\ }\href {\doibase 10.1126/science.1261422}
  {\bibfield  {journal} {\bibinfo  {journal} {Science}\ }\textbf {\bibinfo
  {volume} {348}},\ \bibinfo {pages} {1115--1118} (\bibinfo {year}
  {2015})}\BibitemShut {NoStop}%
\bibitem [{\citenamefont {Bylinskii}\ \emph {et~al.}(2016)\citenamefont
  {Bylinskii}, \citenamefont {Gangloff}, \citenamefont {Counts},\ and\
  \citenamefont {Vuleti{\'c}}}]{Bylinskii:2016}%
  \BibitemOpen
  \bibfield  {author} {\bibinfo {author} {\bibfnamefont {A.}~\bibnamefont
  {Bylinskii}}, \bibinfo {author} {\bibfnamefont {D.}~\bibnamefont {Gangloff}},
  \bibinfo {author} {\bibfnamefont {I.}~\bibnamefont {Counts}}, \ and\ \bibinfo
  {author} {\bibfnamefont {V.}~\bibnamefont {Vuleti{\'c}}},\ }\bibfield
  {title} {\enquote {\bibinfo {title} {Observation of aubry-type transition in
  finite atom chains via friction},}\ }\href {\doibase 10.1038/nmat4601}
  {\bibfield  {journal} {\bibinfo  {journal} {Nature Materials}\ }\textbf
  {\bibinfo {volume} {15}},\ \bibinfo {pages} {717--721} (\bibinfo {year}
  {2016})}\BibitemShut {NoStop}%
\bibitem [{\citenamefont {Gangloff}\ \emph {et~al.}(2020)\citenamefont
  {Gangloff}, \citenamefont {Bylinskii},\ and\ \citenamefont
  {Vuleti\ifmmode~\acute{c}\else \'{c}\fi{}}}]{Gangloff:2022}%
  \BibitemOpen
  \bibfield  {author} {\bibinfo {author} {\bibfnamefont {D.~A.}\ \bibnamefont
  {Gangloff}}, \bibinfo {author} {\bibfnamefont {A.}~\bibnamefont {Bylinskii}},
  \ and\ \bibinfo {author} {\bibfnamefont {V.}~\bibnamefont
  {Vuleti\ifmmode~\acute{c}\else \'{c}\fi{}}},\ }\bibfield  {title} {\enquote
  {\bibinfo {title} {Kinks and nanofriction: Structural phases in few-atom
  chains},}\ }\href {\doibase 10.1103/PhysRevResearch.2.013380} {\bibfield
  {journal} {\bibinfo  {journal} {Phys. Rev. Research}\ }\textbf {\bibinfo
  {volume} {2}},\ \bibinfo {pages} {013380} (\bibinfo {year}
  {2020})}\BibitemShut {NoStop}%
\bibitem [{\citenamefont {Kiethe}\ \emph {et~al.}(2017)\citenamefont {Kiethe},
  \citenamefont {Nigmatullin}, \citenamefont {Kalincev}, \citenamefont
  {Schmirander},\ and\ \citenamefont {Mehlst\"aubler}}]{Kiethe:2017}%
  \BibitemOpen
  \bibfield  {author} {\bibinfo {author} {\bibfnamefont {J.}~\bibnamefont
  {Kiethe}}, \bibinfo {author} {\bibfnamefont {R.}~\bibnamefont {Nigmatullin}},
  \bibinfo {author} {\bibfnamefont {D.}~\bibnamefont {Kalincev}}, \bibinfo
  {author} {\bibfnamefont {T.}~\bibnamefont {Schmirander}}, \ and\ \bibinfo
  {author} {\bibfnamefont {T.E.}\ \bibnamefont {Mehlst\"aubler}},\ }\bibfield
  {title} {\enquote {\bibinfo {title} {Probing nanofriction and aubry-type
  signatures in a finite self-organized system},}\ }\href@noop {} {\bibfield
  {journal} {\bibinfo  {journal} {Nature Commun.}\ }\textbf {\bibinfo {volume}
  {8}},\ \bibinfo {pages} {15364} (\bibinfo {year} {2017})}\BibitemShut
  {NoStop}%
\bibitem [{\citenamefont {Kiethe}\ \emph {et~al.}(2018)\citenamefont {Kiethe},
  \citenamefont {Nigmatullin}, \citenamefont {Schmirander}, \citenamefont
  {Kalincev},\ and\ \citenamefont {Mehlst\"aubler}}]{Kiethe:2018}%
  \BibitemOpen
  \bibfield  {author} {\bibinfo {author} {\bibfnamefont {J.}~\bibnamefont
  {Kiethe}}, \bibinfo {author} {\bibfnamefont {R.}~\bibnamefont {Nigmatullin}},
  \bibinfo {author} {\bibfnamefont {T.}~\bibnamefont {Schmirander}}, \bibinfo
  {author} {\bibfnamefont {D.}~\bibnamefont {Kalincev}}, \ and\ \bibinfo
  {author} {\bibfnamefont {T.~E.}\ \bibnamefont {Mehlst\"aubler}},\ }\bibfield
  {title} {\enquote {\bibinfo {title} {Nanofriction and motion of topological
  defects in self-organized ion coulomb crystals},}\ }\href {\doibase
  10.1088/1367-2630/aaf3d5} {\bibfield  {journal} {\bibinfo  {journal} {New
  Journal of Physics}\ }\textbf {\bibinfo {volume} {20}},\ \bibinfo {pages}
  {123017} (\bibinfo {year} {2018})}\BibitemShut {NoStop}%
\bibitem [{\citenamefont {Zanca}\ \emph {et~al.}(2018)\citenamefont {Zanca},
  \citenamefont {Pellegrini}, \citenamefont {Santoro},\ and\ \citenamefont
  {Tosatti}}]{TosattiPNAS}%
  \BibitemOpen
  \bibfield  {author} {\bibinfo {author} {\bibfnamefont {T.}~\bibnamefont
  {Zanca}}, \bibinfo {author} {\bibfnamefont {F.}~\bibnamefont {Pellegrini}},
  \bibinfo {author} {\bibfnamefont {G.~E.}\ \bibnamefont {Santoro}}, \ and\
  \bibinfo {author} {\bibfnamefont {E.}~\bibnamefont {Tosatti}},\ }\bibfield
  {title} {\enquote {\bibinfo {title} {Frictional lubricity enhanced by quantum
  mechanics},}\ }\href {\doibase 10.1073/pnas.1801144115} {\bibfield  {journal}
  {\bibinfo  {journal} {Proceedings of the National Academy of Sciences}\
  }\textbf {\bibinfo {volume} {115}},\ \bibinfo {pages} {3547--3550} (\bibinfo
  {year} {2018})}\BibitemShut {NoStop}%
\bibitem [{\citenamefont {Krajewski}\ and\ \citenamefont
  {M\"user}(2004)}]{Mueser:2004}%
  \BibitemOpen
  \bibfield  {author} {\bibinfo {author} {\bibfnamefont {F.~R.}\ \bibnamefont
  {Krajewski}}\ and\ \bibinfo {author} {\bibfnamefont {M.~H.}\ \bibnamefont
  {M\"user}},\ }\bibfield  {title} {\enquote {\bibinfo {title} {Quantum creep
  and quantum-creep transitions in 1d sine-gordon chains},}\ }\href {\doibase
  10.1103/PhysRevLett.92.030601} {\bibfield  {journal} {\bibinfo  {journal}
  {Phys. Rev. Lett.}\ }\textbf {\bibinfo {volume} {92}},\ \bibinfo {pages}
  {030601} (\bibinfo {year} {2004})}\BibitemShut {NoStop}%
\bibitem [{\citenamefont {Krajewski}\ and\ \citenamefont
  {M{\"u}ser}(2005)}]{Mueser:2005}%
  \BibitemOpen
  \bibfield  {author} {\bibinfo {author} {\bibfnamefont {F.~R.}\ \bibnamefont
  {Krajewski}}\ and\ \bibinfo {author} {\bibfnamefont {M.~H.}\ \bibnamefont
  {M{\"u}ser}},\ }\bibfield  {title} {\enquote {\bibinfo {title} {Quantum
  dynamics in the highly discrete, commensurate frenkel kontorova model: a
  path-integral molecular dynamics study.}}\ }\href@noop {} {\bibfield
  {journal} {\bibinfo  {journal} {The Journal of chemical physics}\ }\textbf
  {\bibinfo {volume} {122 12}},\ \bibinfo {pages} {124711} (\bibinfo {year}
  {2005})}\BibitemShut {NoStop}%
\bibitem [{\citenamefont {Bonetti}\ \emph {et~al.}(2021)\citenamefont
  {Bonetti}, \citenamefont {Rucci}, \citenamefont {Chiofalo},\ and\
  \citenamefont {Vuleti\'c}}]{Bonetti}%
  \BibitemOpen
  \bibfield  {author} {\bibinfo {author} {\bibfnamefont {P.~M.}\ \bibnamefont
  {Bonetti}}, \bibinfo {author} {\bibfnamefont {A.}~\bibnamefont {Rucci}},
  \bibinfo {author} {\bibfnamefont {M.~L.}\ \bibnamefont {Chiofalo}}, \ and\
  \bibinfo {author} {\bibfnamefont {V.}~\bibnamefont {Vuleti\'c}},\ }\bibfield
  {title} {\enquote {\bibinfo {title} {Quantum effects in the aubry
  transition},}\ }\href {\doibase 10.1103/PhysRevResearch.3.013031} {\bibfield
  {journal} {\bibinfo  {journal} {Phys. Rev. Res.}\ }\textbf {\bibinfo {volume}
  {3}},\ \bibinfo {pages} {013031} (\bibinfo {year} {2021})}\BibitemShut
  {NoStop}%
\bibitem [{\citenamefont {Timm}\ \emph {et~al.}(2021)\citenamefont {Timm},
  \citenamefont {R\"uffert}, \citenamefont {Weimer}, \citenamefont {Santos},\
  and\ \citenamefont {Mehlst\"aubler}}]{Timm:2021}%
  \BibitemOpen
  \bibfield  {author} {\bibinfo {author} {\bibfnamefont {L.}~\bibnamefont
  {Timm}}, \bibinfo {author} {\bibfnamefont {L.~A.}\ \bibnamefont {R\"uffert}},
  \bibinfo {author} {\bibfnamefont {H.}~\bibnamefont {Weimer}}, \bibinfo
  {author} {\bibfnamefont {L.}~\bibnamefont {Santos}}, \ and\ \bibinfo {author}
  {\bibfnamefont {T.~E.}\ \bibnamefont {Mehlst\"aubler}},\ }\bibfield  {title}
  {\enquote {\bibinfo {title} {Quantum nanofriction in trapped ion chains with
  a topological defect},}\ }\href {\doibase 10.1103/PhysRevResearch.3.043141}
  {\bibfield  {journal} {\bibinfo  {journal} {Phys. Rev. Res.}\ }\textbf
  {\bibinfo {volume} {3}},\ \bibinfo {pages} {043141} (\bibinfo {year}
  {2021})}\BibitemShut {NoStop}%
\bibitem [{\citenamefont {Chelpanova}\ \emph {et~al.}(2023)\citenamefont
  {Chelpanova}, \citenamefont {Kelly}, \citenamefont {Morigi}, \citenamefont
  {Schmidt-Kaler},\ and\ \citenamefont {Marino}}]{Chelpanova:2023}%
  \BibitemOpen
  \bibfield  {author} {\bibinfo {author} {\bibfnamefont {O.}~\bibnamefont
  {Chelpanova}}, \bibinfo {author} {\bibfnamefont {S.~P.}\ \bibnamefont
  {Kelly}}, \bibinfo {author} {\bibfnamefont {G.}~\bibnamefont {Morigi}},
  \bibinfo {author} {\bibfnamefont {F.}~\bibnamefont {Schmidt-Kaler}}, \ and\
  \bibinfo {author} {\bibfnamefont {J.}~\bibnamefont {Marino}},\ }\bibfield
  {title} {\enquote {\bibinfo {title} {Injection and nucleation of topological
  defects in the quench dynamics of the frenkel-kontorova model},}\ }\href
  {\doibase 10.1209/0295-5075/ace27d} {\bibfield  {journal} {\bibinfo
  {journal} {EPL}\ }\textbf {\bibinfo {volume} {143}},\ \bibinfo {pages}
  {25002} (\bibinfo {year} {2023})}\BibitemShut {NoStop}%
\bibitem [{\citenamefont {Schulz}(1993)}]{Schulz:1993}%
  \BibitemOpen
  \bibfield  {author} {\bibinfo {author} {\bibfnamefont {H.~J.}\ \bibnamefont
  {Schulz}},\ }\bibfield  {title} {\enquote {\bibinfo {title} {Wigner crystal
  in one dimension},}\ }\href {\doibase 10.1103/PhysRevLett.71.1864} {\bibfield
   {journal} {\bibinfo  {journal} {Phys. Rev. Lett.}\ }\textbf {\bibinfo
  {volume} {71}},\ \bibinfo {pages} {1864--1867} (\bibinfo {year}
  {1993})}\BibitemShut {NoStop}%
\bibitem [{\citenamefont {Landa}\ \emph {et~al.}(2020)\citenamefont {Landa},
  \citenamefont {Cormick},\ and\ \citenamefont {Morigi}}]{Landa:2020}%
  \BibitemOpen
  \bibfield  {author} {\bibinfo {author} {\bibfnamefont {H.}~\bibnamefont
  {Landa}}, \bibinfo {author} {\bibfnamefont {C.}~\bibnamefont {Cormick}}, \
  and\ \bibinfo {author} {\bibfnamefont {G.}~\bibnamefont {Morigi}},\
  }\bibfield  {title} {\enquote {\bibinfo {title} {Static kinks in chains of
  interacting atoms},}\ }\href {\doibase 10.3390/condmat5020035} {\bibfield
  {journal} {\bibinfo  {journal} {Condensed Matter}\ }\textbf {\bibinfo
  {volume} {5}} (\bibinfo {year} {2020}),\ 10.3390/condmat5020035}\BibitemShut
  {NoStop}%
\bibitem [{\citenamefont {Fishman}\ \emph {et~al.}(2008)\citenamefont
  {Fishman}, \citenamefont {De~Chiara}, \citenamefont {Calarco},\ and\
  \citenamefont {Morigi}}]{Fishman:2008}%
  \BibitemOpen
  \bibfield  {author} {\bibinfo {author} {\bibfnamefont {S.}~\bibnamefont
  {Fishman}}, \bibinfo {author} {\bibfnamefont {G.}~\bibnamefont {De~Chiara}},
  \bibinfo {author} {\bibfnamefont {T.}~\bibnamefont {Calarco}}, \ and\
  \bibinfo {author} {\bibfnamefont {G.}~\bibnamefont {Morigi}},\ }\bibfield
  {title} {\enquote {\bibinfo {title} {Structural phase transitions in
  low-dimensional ion crystals},}\ }\href {\doibase 10.1103/PhysRevB.77.064111}
  {\bibfield  {journal} {\bibinfo  {journal} {Phys. Rev. B}\ }\textbf {\bibinfo
  {volume} {77}},\ \bibinfo {pages} {064111} (\bibinfo {year}
  {2008})}\BibitemShut {NoStop}%
\bibitem [{\citenamefont {Pokrovsky}\ and\ \citenamefont
  {Virosztek}(1983)}]{Pokrovsky_1983}%
  \BibitemOpen
  \bibfield  {author} {\bibinfo {author} {\bibfnamefont {V.L.}\ \bibnamefont
  {Pokrovsky}}\ and\ \bibinfo {author} {\bibfnamefont {A.}~\bibnamefont
  {Virosztek}},\ }\bibfield  {title} {\enquote {\bibinfo {title} {Long-range
  interactions in commensurate-incommensurate phase transition},}\ }\href
  {\doibase 10.1088/0022-3719/16/23/013} {\bibfield  {journal} {\bibinfo
  {journal} {Journal of Physics C: Solid State Physics}\ }\textbf {\bibinfo
  {volume} {16}},\ \bibinfo {pages} {4513--4525} (\bibinfo {year}
  {1983})}\BibitemShut {NoStop}%
\bibitem [{\citenamefont {Braun}\ \emph {et~al.}(1990)\citenamefont {Braun},
  \citenamefont {Kivshar},\ and\ \citenamefont {Zelenskaya}}]{Braun:1990}%
  \BibitemOpen
  \bibfield  {author} {\bibinfo {author} {\bibfnamefont {O.~M.}\ \bibnamefont
  {Braun}}, \bibinfo {author} {\bibfnamefont {Yu.~S.}\ \bibnamefont {Kivshar}},
  \ and\ \bibinfo {author} {\bibfnamefont {I.~I.}\ \bibnamefont {Zelenskaya}},\
  }\bibfield  {title} {\enquote {\bibinfo {title} {Kinks in the
  frenkel-kontorova model with long-range interparticle interactions},}\ }\href
  {\doibase 10.1103/PhysRevB.41.7118} {\bibfield  {journal} {\bibinfo
  {journal} {Phys. Rev. B}\ }\textbf {\bibinfo {volume} {41}},\ \bibinfo
  {pages} {7118--7138} (\bibinfo {year} {1990})}\BibitemShut {NoStop}%
\bibitem [{SM()}]{SM}%
  \BibitemOpen
  \bibinfo {note} {See Supplemental Material for (A) the full derivation of the
  (1+1) Thirring model with long-range interactions starting from the
  long-range Frenkel-Kontorova model, (B) the lattice field theoretical
  description of the Thirring model and its mapping onto a spin model, (C) the
  discussion of the Kac scaling of long-range interactions and the mean-field
  limit. The Supplementary Materials include the references
  \cite{CAMPA200957,Morigi:2004,Mandelstam:1975,Defenu:2023,Morigi:2004L}.}\BibitemShut
  {Stop}%
\bibitem [{\citenamefont {Thirring}(1958)}]{Thirring:1958}%
  \BibitemOpen
  \bibfield  {author} {\bibinfo {author} {\bibfnamefont {W.E.}\ \bibnamefont
  {Thirring}},\ }\bibfield  {title} {\enquote {\bibinfo {title} {A soluble
  relativistic field theory},}\ }\href {\doibase
  https://doi.org/10.1016/0003-4916(58)90015-0} {\bibfield  {journal} {\bibinfo
   {journal} {Annals of Physics}\ }\textbf {\bibinfo {volume} {3}},\ \bibinfo
  {pages} {91--112} (\bibinfo {year} {1958})}\BibitemShut {NoStop}%
\bibitem [{\citenamefont {Coleman}(1975)}]{Coleman}%
  \BibitemOpen
  \bibfield  {author} {\bibinfo {author} {\bibfnamefont {S.}~\bibnamefont
  {Coleman}},\ }\bibfield  {title} {\enquote {\bibinfo {title} {Quantum
  sine-gordon equation as the massive thirring model},}\ }\href {\doibase
  10.1103/PhysRevD.11.2088} {\bibfield  {journal} {\bibinfo  {journal} {Phys.
  Rev. D}\ }\textbf {\bibinfo {volume} {11}},\ \bibinfo {pages} {2088--2097}
  (\bibinfo {year} {1975})}\BibitemShut {NoStop}%
\bibitem [{\citenamefont {Mandelstam}(1975)}]{Mandelstam:1975}%
  \BibitemOpen
  \bibfield  {author} {\bibinfo {author} {\bibfnamefont {S.}~\bibnamefont
  {Mandelstam}},\ }\bibfield  {title} {\enquote {\bibinfo {title} {Soliton
  operators for the quantized sine-gordon equation},}\ }\href {\doibase
  10.1103/PhysRevD.11.3026} {\bibfield  {journal} {\bibinfo  {journal} {Phys.
  Rev. D}\ }\textbf {\bibinfo {volume} {11}},\ \bibinfo {pages} {3026--3030}
  (\bibinfo {year} {1975})}\BibitemShut {NoStop}%
\bibitem [{\citenamefont {Ba{\~{n}}uls}\ \emph {et~al.}(2020)\citenamefont
  {Ba{\~{n}}uls}, \citenamefont {Blatt}, \citenamefont {Catani}, \citenamefont
  {Celi}, \citenamefont {Cirac}, \citenamefont {Dalmonte}, \citenamefont
  {Fallani}, \citenamefont {Jansen}, \citenamefont {Lewenstein}, \citenamefont
  {Montangero}, \citenamefont {Muschik}, \citenamefont {Reznik}, \citenamefont
  {Rico}, \citenamefont {Tagliacozzo}, \citenamefont {Van~Acoleyen},
  \citenamefont {Verstraete}, \citenamefont {Wiese}, \citenamefont {Wingate},
  \citenamefont {Zakrzewski},\ and\ \citenamefont {Zoller}}]{Banuls_2020}%
  \BibitemOpen
  \bibfield  {author} {\bibinfo {author} {\bibfnamefont {M.~C.}\ \bibnamefont
  {Ba{\~{n}}uls}}, \bibinfo {author} {\bibfnamefont {R.}~\bibnamefont {Blatt}},
  \bibinfo {author} {\bibfnamefont {J.}~\bibnamefont {Catani}}, \bibinfo
  {author} {\bibfnamefont {A.}~\bibnamefont {Celi}}, \bibinfo {author}
  {\bibfnamefont {J.~I.}\ \bibnamefont {Cirac}}, \bibinfo {author}
  {\bibfnamefont {M.}~\bibnamefont {Dalmonte}}, \bibinfo {author}
  {\bibfnamefont {L.}~\bibnamefont {Fallani}}, \bibinfo {author} {\bibfnamefont
  {K.}~\bibnamefont {Jansen}}, \bibinfo {author} {\bibfnamefont
  {M.}~\bibnamefont {Lewenstein}}, \bibinfo {author} {\bibfnamefont
  {S.}~\bibnamefont {Montangero}}, \bibinfo {author} {\bibfnamefont {C.~A.}\
  \bibnamefont {Muschik}}, \bibinfo {author} {\bibfnamefont {B.}~\bibnamefont
  {Reznik}}, \bibinfo {author} {\bibfnamefont {E.}~\bibnamefont {Rico}},
  \bibinfo {author} {\bibfnamefont {L.}~\bibnamefont {Tagliacozzo}}, \bibinfo
  {author} {\bibfnamefont {K.}~\bibnamefont {Van~Acoleyen}}, \bibinfo {author}
  {\bibfnamefont {F.}~\bibnamefont {Verstraete}}, \bibinfo {author}
  {\bibfnamefont {U.-J.}\ \bibnamefont {Wiese}}, \bibinfo {author}
  {\bibfnamefont {M.}~\bibnamefont {Wingate}}, \bibinfo {author} {\bibfnamefont
  {J.}~\bibnamefont {Zakrzewski}}, \ and\ \bibinfo {author} {\bibfnamefont
  {P.}~\bibnamefont {Zoller}},\ }\bibfield  {title} {\enquote {\bibinfo {title}
  {Simulating lattice gauge theories within quantum technologies},}\ }\href
  {\doibase 10.1140/epjd/e2020-100571-8} {\bibfield  {journal} {\bibinfo
  {journal} {The European Physical Journal D}\ }\textbf {\bibinfo {volume}
  {74}},\ \bibinfo {pages} {165} (\bibinfo {year} {2020})}\BibitemShut
  {NoStop}%
\bibitem [{Note1()}]{Note1}%
  \BibitemOpen
  \bibinfo {note} {The action can be generalized to the case $d_0/a=n_0/m_0$,
  with $m_0>1$ and $n_0$ and $m_0$ prime to each other, see e.g.\ \cite
  {Pokrovskij_Talapov1}. Here, we restrict to the case $m_0=1$.}\BibitemShut
  {Stop}%
\bibitem [{\citenamefont {Susskind}(1977)}]{Susskind:1977}%
  \BibitemOpen
  \bibfield  {author} {\bibinfo {author} {\bibfnamefont {L.}~\bibnamefont
  {Susskind}},\ }\bibfield  {title} {\enquote {\bibinfo {title} {Lattice
  fermions},}\ }\href {\doibase 10.1103/PhysRevD.16.3031} {\bibfield  {journal}
  {\bibinfo  {journal} {Phys. Rev. D}\ }\textbf {\bibinfo {volume} {16}},\
  \bibinfo {pages} {3031--3039} (\bibinfo {year} {1977})}\BibitemShut {NoStop}%
\bibitem [{\citenamefont {Ba\~nuls}\ \emph {et~al.}(2019)\citenamefont
  {Ba\~nuls}, \citenamefont {Cichy}, \citenamefont {Kao}, \citenamefont {Lin},
  \citenamefont {Lin},\ and\ \citenamefont {Tan}}]{Banuls}%
  \BibitemOpen
  \bibfield  {author} {\bibinfo {author} {\bibfnamefont {M.~C.}\ \bibnamefont
  {Ba\~nuls}}, \bibinfo {author} {\bibfnamefont {K.}~\bibnamefont {Cichy}},
  \bibinfo {author} {\bibfnamefont {Y.-J.}\ \bibnamefont {Kao}}, \bibinfo
  {author} {\bibfnamefont {C.-J.~D.}\ \bibnamefont {Lin}}, \bibinfo {author}
  {\bibfnamefont {Y.-P.}\ \bibnamefont {Lin}}, \ and\ \bibinfo {author}
  {\bibfnamefont {D.~T.-L.}\ \bibnamefont {Tan}},\ }\bibfield  {title}
  {\enquote {\bibinfo {title} {Phase structure of the ($1+1$)-dimensional
  massive thirring model from matrix product states},}\ }\href {\doibase
  10.1103/PhysRevD.100.094504} {\bibfield  {journal} {\bibinfo  {journal}
  {Phys. Rev. D}\ }\textbf {\bibinfo {volume} {100}},\ \bibinfo {pages}
  {094504} (\bibinfo {year} {2019})}\BibitemShut {NoStop}%
\bibitem [{\citenamefont {Morigi}\ and\ \citenamefont
  {Fishman}(2004{\natexlab{a}})}]{Morigi:2004}%
  \BibitemOpen
  \bibfield  {author} {\bibinfo {author} {\bibfnamefont {G.}~\bibnamefont
  {Morigi}}\ and\ \bibinfo {author} {\bibfnamefont {S.}~\bibnamefont
  {Fishman}},\ }\bibfield  {title} {\enquote {\bibinfo {title} {Dynamics of an
  ion chain in a harmonic potential},}\ }\href {\doibase
  10.1103/PhysRevE.70.066141} {\bibfield  {journal} {\bibinfo  {journal} {Phys.
  Rev. E}\ }\textbf {\bibinfo {volume} {70}},\ \bibinfo {pages} {066141}
  (\bibinfo {year} {2004}{\natexlab{a}})}\BibitemShut {NoStop}%
\bibitem [{\citenamefont {Koziol}\ \emph {et~al.}(2023)\citenamefont {Koziol},
  \citenamefont {Duft}, \citenamefont {Morigi},\ and\ \citenamefont
  {Schmidt}}]{Koziol:2023}%
  \BibitemOpen
  \bibfield  {author} {\bibinfo {author} {\bibfnamefont {J.~A.}\ \bibnamefont
  {Koziol}}, \bibinfo {author} {\bibfnamefont {A.}~\bibnamefont {Duft}},
  \bibinfo {author} {\bibfnamefont {G.}~\bibnamefont {Morigi}}, \ and\ \bibinfo
  {author} {\bibfnamefont {K.~P.}\ \bibnamefont {Schmidt}},\ }\bibfield
  {title} {\enquote {\bibinfo {title} {{Systematic analysis of crystalline
  phases in bosonic lattice models with algebraically decaying density-density
  interactions}},}\ }\href {\doibase 10.21468/SciPostPhys.14.5.136} {\bibfield
  {journal} {\bibinfo  {journal} {SciPost Phys.}\ }\textbf {\bibinfo {volume}
  {14}},\ \bibinfo {pages} {136} (\bibinfo {year} {2023})}\BibitemShut
  {NoStop}%
\bibitem [{\citenamefont {Biham}\ and\ \citenamefont
  {Mukamel}(1989)}]{Biham:1989}%
  \BibitemOpen
  \bibfield  {author} {\bibinfo {author} {\bibfnamefont {O.}~\bibnamefont
  {Biham}}\ and\ \bibinfo {author} {\bibfnamefont {D.}~\bibnamefont
  {Mukamel}},\ }\bibfield  {title} {\enquote {\bibinfo {title} {Global
  universality in the frenkel-kontorova model},}\ }\href {\doibase
  10.1103/PhysRevA.39.5326} {\bibfield  {journal} {\bibinfo  {journal} {Phys.
  Rev. A}\ }\textbf {\bibinfo {volume} {39}},\ \bibinfo {pages} {5326--5335}
  (\bibinfo {year} {1989})}\BibitemShut {NoStop}%
\bibitem [{\citenamefont {Bak}\ and\ \citenamefont
  {Bruinsma}(1982)}]{Bak-PRL:1982}%
  \BibitemOpen
  \bibfield  {author} {\bibinfo {author} {\bibfnamefont {P.}~\bibnamefont
  {Bak}}\ and\ \bibinfo {author} {\bibfnamefont {R.}~\bibnamefont {Bruinsma}},\
  }\bibfield  {title} {\enquote {\bibinfo {title} {One-dimensional ising model
  and the complete devil's staircase},}\ }\href {\doibase
  10.1103/PhysRevLett.49.249} {\bibfield  {journal} {\bibinfo  {journal} {Phys.
  Rev. Lett.}\ }\textbf {\bibinfo {volume} {49}},\ \bibinfo {pages} {249--251}
  (\bibinfo {year} {1982})}\BibitemShut {NoStop}%
\bibitem [{\citenamefont {Bruinsma}\ and\ \citenamefont
  {Aeppli}(1983)}]{Bruisma:1983a}%
  \BibitemOpen
  \bibfield  {author} {\bibinfo {author} {\bibfnamefont {R.}~\bibnamefont
  {Bruinsma}}\ and\ \bibinfo {author} {\bibfnamefont {G.}~\bibnamefont
  {Aeppli}},\ }\bibfield  {title} {\enquote {\bibinfo {title} {One-dimensional
  ising model in a random field},}\ }\href {\doibase
  10.1103/PhysRevLett.50.1494} {\bibfield  {journal} {\bibinfo  {journal}
  {Phys. Rev. Lett.}\ }\textbf {\bibinfo {volume} {50}},\ \bibinfo {pages}
  {1494--1497} (\bibinfo {year} {1983})}\BibitemShut {NoStop}%
\bibitem [{\citenamefont {Bruinsma}\ and\ \citenamefont
  {Bak}(1983)}]{Bruinsma:1983}%
  \BibitemOpen
  \bibfield  {author} {\bibinfo {author} {\bibfnamefont {R.}~\bibnamefont
  {Bruinsma}}\ and\ \bibinfo {author} {\bibfnamefont {P.}~\bibnamefont {Bak}},\
  }\bibfield  {title} {\enquote {\bibinfo {title} {Self-similarity and fractal
  dimension of the devil's staircase in the one-dimensional ising model},}\
  }\href {\doibase 10.1103/PhysRevB.27.5824} {\bibfield  {journal} {\bibinfo
  {journal} {Phys. Rev. B}\ }\textbf {\bibinfo {volume} {27}},\ \bibinfo
  {pages} {5824--5825} (\bibinfo {year} {1983})}\BibitemShut {NoStop}%
\bibitem [{\citenamefont {Menu}\ \emph {et~al.}(2024)\citenamefont {Menu},
  \citenamefont {Malo}, \citenamefont {Vuletić}, \citenamefont {Chiofalo},\
  and\ \citenamefont {Morigi}}]{Menu:2024}%
  \BibitemOpen
  \bibfield  {author} {\bibinfo {author} {\bibfnamefont {R.}~\bibnamefont
  {Menu}}, \bibinfo {author} {\bibfnamefont {J.~Yago}\ \bibnamefont {Malo}},
  \bibinfo {author} {\bibfnamefont {V.}~\bibnamefont {Vuletić}}, \bibinfo
  {author} {\bibfnamefont {M.~L;}\ \bibnamefont {Chiofalo}}, \ and\ \bibinfo
  {author} {\bibfnamefont {G;}~\bibnamefont {Morigi}},\ }\href
  {https://arxiv.org/abs/2403.15843} {\enquote {\bibinfo {title} {Fractal
  ground state of ion chains in periodic potentials},}\ } (\bibinfo {year}
  {2024}),\ \Eprint {http://arxiv.org/abs/2403.15843} {arXiv:2403.15843
  [cond-mat.quant-gas]} \BibitemShut {NoStop}%
\bibitem [{\citenamefont {Li}\ \emph {et~al.}(2017)\citenamefont {Li},
  \citenamefont {Urban}, \citenamefont {Noel}, \citenamefont {Chuang},
  \citenamefont {Xia}, \citenamefont {Ransford}, \citenamefont {Hemmerling},
  \citenamefont {Wang}, \citenamefont {Li}, \citenamefont {H\"affner},\ and\
  \citenamefont {Zhang}}]{Haeffner}%
  \BibitemOpen
  \bibfield  {author} {\bibinfo {author} {\bibfnamefont {H.-K.}\ \bibnamefont
  {Li}}, \bibinfo {author} {\bibfnamefont {E.}~\bibnamefont {Urban}}, \bibinfo
  {author} {\bibfnamefont {C.}~\bibnamefont {Noel}}, \bibinfo {author}
  {\bibfnamefont {A.}~\bibnamefont {Chuang}}, \bibinfo {author} {\bibfnamefont
  {Y.}~\bibnamefont {Xia}}, \bibinfo {author} {\bibfnamefont {A.}~\bibnamefont
  {Ransford}}, \bibinfo {author} {\bibfnamefont {B.}~\bibnamefont
  {Hemmerling}}, \bibinfo {author} {\bibfnamefont {Y.}~\bibnamefont {Wang}},
  \bibinfo {author} {\bibfnamefont {T.}~\bibnamefont {Li}}, \bibinfo {author}
  {\bibfnamefont {H.}~\bibnamefont {H\"affner}}, \ and\ \bibinfo {author}
  {\bibfnamefont {X.}~\bibnamefont {Zhang}},\ }\bibfield  {title} {\enquote
  {\bibinfo {title} {Realization of translational symmetry in trapped cold ion
  rings},}\ }\href {\doibase 10.1103/PhysRevLett.118.053001} {\bibfield
  {journal} {\bibinfo  {journal} {Phys. Rev. Lett.}\ }\textbf {\bibinfo
  {volume} {118}},\ \bibinfo {pages} {053001} (\bibinfo {year}
  {2017})}\BibitemShut {NoStop}%
\bibitem [{\citenamefont {Dubin}(1997)}]{Dubin:1997}%
  \BibitemOpen
  \bibfield  {author} {\bibinfo {author} {\bibfnamefont {D.~H.~E.}\
  \bibnamefont {Dubin}},\ }\bibfield  {title} {\enquote {\bibinfo {title}
  {Minimum energy state of the one-dimensional coulomb chain},}\ }\href
  {\doibase 10.1103/PhysRevE.55.4017} {\bibfield  {journal} {\bibinfo
  {journal} {Phys. Rev. E}\ }\textbf {\bibinfo {volume} {55}},\ \bibinfo
  {pages} {4017--4028} (\bibinfo {year} {1997})}\BibitemShut {NoStop}%
\bibitem [{\citenamefont {Fogarty}\ \emph {et~al.}(2015)\citenamefont
  {Fogarty}, \citenamefont {Cormick}, \citenamefont {Landa}, \citenamefont
  {Stojanovi\ifmmode~\acute{c}\else \'{c}\fi{}}, \citenamefont {Demler},\ and\
  \citenamefont {Morigi}}]{Fogarty:2015}%
  \BibitemOpen
  \bibfield  {author} {\bibinfo {author} {\bibfnamefont {T.}~\bibnamefont
  {Fogarty}}, \bibinfo {author} {\bibfnamefont {C.}~\bibnamefont {Cormick}},
  \bibinfo {author} {\bibfnamefont {H.}~\bibnamefont {Landa}}, \bibinfo
  {author} {\bibfnamefont {Vladimir~M.}\ \bibnamefont
  {Stojanovi\ifmmode~\acute{c}\else \'{c}\fi{}}}, \bibinfo {author}
  {\bibfnamefont {E.}~\bibnamefont {Demler}}, \ and\ \bibinfo {author}
  {\bibfnamefont {Giovanna}\ \bibnamefont {Morigi}},\ }\bibfield  {title}
  {\enquote {\bibinfo {title} {Nanofriction in cavity quantum
  electrodynamics},}\ }\href {\doibase 10.1103/PhysRevLett.115.233602}
  {\bibfield  {journal} {\bibinfo  {journal} {Phys. Rev. Lett.}\ }\textbf
  {\bibinfo {volume} {115}},\ \bibinfo {pages} {233602} (\bibinfo {year}
  {2015})}\BibitemShut {NoStop}%
\bibitem [{\citenamefont {Rotondo}\ \emph {et~al.}(2016)\citenamefont
  {Rotondo}, \citenamefont {Molinari}, \citenamefont {Ratti},\ and\
  \citenamefont {Gherardi}}]{Rotondo_2016}%
  \BibitemOpen
  \bibfield  {author} {\bibinfo {author} {\bibfnamefont {P.}~\bibnamefont
  {Rotondo}}, \bibinfo {author} {\bibfnamefont {L.~G.}\ \bibnamefont
  {Molinari}}, \bibinfo {author} {\bibfnamefont {P.}~\bibnamefont {Ratti}}, \
  and\ \bibinfo {author} {\bibfnamefont {M.}~\bibnamefont {Gherardi}},\
  }\bibfield  {title} {\enquote {\bibinfo {title} {Devil's staircase phase
  diagram of the fractional quantum hall effect in the thin-torus limit},}\
  }\href {\doibase 10.1103/PhysRevLett.116.256803} {\bibfield  {journal}
  {\bibinfo  {journal} {Phys. Rev. Lett.}\ }\textbf {\bibinfo {volume} {116}},\
  \bibinfo {pages} {256803} (\bibinfo {year} {2016})}\BibitemShut {NoStop}%
\bibitem [{\citenamefont {Sagi}\ and\ \citenamefont
  {Nussinov}(2016)}]{Nussinov:2016}%
  \BibitemOpen
  \bibfield  {author} {\bibinfo {author} {\bibfnamefont {E.}~\bibnamefont
  {Sagi}}\ and\ \bibinfo {author} {\bibfnamefont {Z.}~\bibnamefont
  {Nussinov}},\ }\bibfield  {title} {\enquote {\bibinfo {title} {Emergent
  quasicrystals in strongly correlated systems},}\ }\href {\doibase
  10.1103/PhysRevB.94.035131} {\bibfield  {journal} {\bibinfo  {journal} {Phys.
  Rev. B}\ }\textbf {\bibinfo {volume} {94}},\ \bibinfo {pages} {035131}
  (\bibinfo {year} {2016})}\BibitemShut {NoStop}%
\bibitem [{\citenamefont {Petrova}\ \emph {et~al.}(2024)\citenamefont
  {Petrova}, \citenamefont {Tiunov}, \citenamefont {Ba\~nuls},\ and\
  \citenamefont {Fedorov}}]{Petrova:2024}%
  \BibitemOpen
  \bibfield  {author} {\bibinfo {author} {\bibfnamefont {E.~V.}\ \bibnamefont
  {Petrova}}, \bibinfo {author} {\bibfnamefont {E.~S.}\ \bibnamefont {Tiunov}},
  \bibinfo {author} {\bibfnamefont {M.~C.}\ \bibnamefont {Ba\~nuls}}, \ and\
  \bibinfo {author} {\bibfnamefont {A.~K.}\ \bibnamefont {Fedorov}},\
  }\bibfield  {title} {\enquote {\bibinfo {title} {Fractal states of the
  schwinger model},}\ }\href {\doibase 10.1103/PhysRevLett.132.050401}
  {\bibfield  {journal} {\bibinfo  {journal} {Phys. Rev. Lett.}\ }\textbf
  {\bibinfo {volume} {132}},\ \bibinfo {pages} {050401} (\bibinfo {year}
  {2024})}\BibitemShut {NoStop}%
\bibitem [{\citenamefont {Morigi}\ and\ \citenamefont
  {Fishman}(2004{\natexlab{b}})}]{Morigi:2004L}%
  \BibitemOpen
  \bibfield  {author} {\bibinfo {author} {\bibfnamefont {G.}~\bibnamefont
  {Morigi}}\ and\ \bibinfo {author} {\bibfnamefont {S.}~\bibnamefont
  {Fishman}},\ }\bibfield  {title} {\enquote {\bibinfo {title} {Eigenmodes and
  thermodynamics of a coulomb chain in a harmonic potential},}\ }\href
  {\doibase 10.1103/PhysRevLett.93.170602} {\bibfield  {journal} {\bibinfo
  {journal} {Phys. Rev. Lett.}\ }\textbf {\bibinfo {volume} {93}},\ \bibinfo
  {pages} {170602} (\bibinfo {year} {2004}{\natexlab{b}})}\BibitemShut
  {NoStop}%
\end{thebibliography}%

\begin{appendix}
\section{Derivation of the (1+1)-Thirring model with long-range interactions}
\label{appA}

Considering $N$ ions of mass $m$ forming a Wigner crystal of stiffness $K$, the local phase shift $\theta_j$ attached to the $j$-th ion follows the the classical equation of motion
\begin{equation*}
    m \ddot{\theta}_j -\sum_{r\neq 0}\dfrac{K}{\vert r\vert^3}(\theta_{j+r}-\theta_j -2\pi\delta r) + \dfrac{(2\pi)^2 V_0}{a^2}\sin\theta=0,
\end{equation*}
where $\delta$ is the discommensuration. In the limit where the phase shifts $\theta_j$ can be treated as one single field $\theta(x,t)$ (where $x$ is expressed in units of $d_0$) whose equation of motion is the one given by Eq.~\eqref{Eq1}:
\begin{align*}
\dfrac{1}{v_s^2} \partial^2_t \theta &=  \partial^2_x\theta - M^2 \sin \theta \notag \\
    &+ \dfrac{1}{3}\partial_x\int_{1}^{N/2}{\dfrac{\partial_x\theta(x+u) + \partial_x\theta(x-u)}{u}\mathrm{d}u},
\end{align*}
with $v_s=\sqrt{\frac{3K}{2m}}$ and $M = \frac{2\pi}{a}\sqrt{\frac{2V_0}{3K}}$.

As for the short-range case, the long-range sine-Gordon model is the Euler-Lagrange equation of the Lagrangian density $\mathcal L$, which gives the action $\mathcal{S}$ of Eq.\ \eqref{Eq2}. Indeed, one recognizes in Eq.~\eqref{Eq1} a d'Alembert operator describing the propagation of the field $\theta(x,t)$, leading to the first part of the action
\begin{equation*}
    \mathcal{S}_1 = \int{\left( \dfrac{1}{2 v_s^2}(\partial_t \theta)^2 - \dfrac{1}{2}(\partial_x\theta)^2\right)\mathrm{d}x}.
\end{equation*}
The non-linear sine term of the Euler-Lagrange equation however stems from a potential leading to a second contribution
\begin{equation*}
    \mathcal{S}_2 = \int M^2 \cos\theta\,  \mathrm{d}x.
\end{equation*}
We recover then the form of the usual sine-Gordon action. The integral term on the other hand results from a third contribution $\mathcal{S}_3=\int\mathcal{L}_3\mathrm{d}x$ such that
\begin{equation*}
    \partial_x\dfrac{\partial\mathcal{L}_3}{\partial(\partial_x \theta)} = \dfrac{1}{3}\partial_x \int_{1}^{N/2}{\dfrac{\partial_x\theta(x+u) + \partial_x\theta(x-u)}{u}\mathrm{d}u}. 
\end{equation*}
The integration of the partial derivative with respect to the variable $\partial_x\theta$ gives rise to a Coulomb-like repulsive interaction between the centers of the solitons, taking the form
\begin{equation*}
    \mathcal{S}_3 = -\dfrac{1}{6}\iint{\dfrac{\partial_x\theta(x)\partial_x\theta(x')}{\vert x-x'\vert}\mathrm{d}x'\mathrm{d}x}.
\end{equation*}
This integral can be split into two contributions via the change of variable $x'=x\pm u$, depending on the position of $x'$ relative to $x$, with $u$ spanning the interval $[1,\,N/2]$.

We note that the discommensuration $\delta$ is absent from the Euler-Lagrange equation and from the total action $\mathcal{S}=\mathcal{S}_1+\mathcal{S}_2+\mathcal{S}_3$. Yet, it is essential to determine the ground state of the system as $\delta$ introduces constraints on the solution of the Euler-Lagrange equation to be energetically favourable. This constraint can be implemented via the prescription $\theta(x) \to \theta(x)-2\pi x\delta$. After rescaling the time $\tau = v_s t$ we recover the action displayed on Eq.~\eqref{Eq2}.

The mapping on the action, Eq.\ \eqref{Eq2}, to the Thirring model is performed by fermionizing the bosonic field $\theta(x,\tau)$ and its conjugate $\dot{\theta}(x,\tau)=\partial_\tau\theta (x,\tau)$. Two fermionic species with different chirality are introduced by the Mandelstam transformation \cite{Mandelstam:1975}:
\begin{subequations}
    \begin{align}
        \psi_1(x) &= \dfrac{1}{\sqrt{2\pi}}\exp\left(-\dfrac{2\pi i}{\beta}\int_{-\infty}^x{\dot{\theta}(u)\mathrm{d}u } - \dfrac{i\beta}{2}\theta(x) \right), \notag\\
        \psi_2(x) &= \dfrac{i}{\sqrt{2\pi}}\exp\left(-\dfrac{2\pi i}{\beta}\int_{-\infty}^x{\dot{\theta}(u)\mathrm{d}u } + \dfrac{i\beta}{2}\theta(x) \right).\notag
    \end{align}
\end{subequations}
Assuming that the soliton field $\theta(x,\tau)$ is slowly varying, then the Baker-Campbell-Hausdorff formula
$$e^A e^B = e^{A+B}e^{[A,B]/2},$$
in the case where the commutator $[A,B]$ is a scalar quantity leads to the relation
\begin{align}
    \psi^\dagger_1(x+1)\psi_1(x)&=-\dfrac{i}{\sqrt{2\pi}}\exp\left[\dfrac{2\pi i}{\beta}\dot{\theta}(x)+\dfrac{i\beta}{2}\theta'(x)\right]\notag\\
    &\simeq -\dfrac{i}{2\pi}+\left(\dfrac{1}{\beta}\dot{\theta}(x)+\dfrac{\beta}{4\pi}\theta'(x) \right)\notag,
\end{align}
where $\theta'(x) = \partial_x\theta(x)$. A renormalization of the total fermion number $\psi^\dagger\psi=\psi^\dagger_1\psi_1 + \psi^\dagger_2 \psi_2$, eliminates the constant term in the previous expression. We use then the approximation $\psi^\dagger_j(x+1)\psi_j(x)=\psi^\dagger_j(x)\psi_j(x)$, to derive the relations
\begin{align*}
    \dfrac{\beta}{2\pi}\theta'(x) &= \psi^\dagger_1(x)\psi_1(x) + \psi^\dagger_2(x)\psi_2(x)\\
     \dfrac{2}{\beta}\dot{\theta}(x) &= \psi^\dagger_1(x)\psi_1(x) - \psi^\dagger_2(x)\psi_2(x)\\
     \cos(\beta\theta(x))&=-\pi (\psi_1^\dagger(x)\psi_2(x) + \psi^\dagger_2(x)\psi_1(x)).
\end{align*}
Additionally, the local density of fermions can be related to kinetic effects via the relations
\begin{align*}
    i\psi^\dagger_1(x)\partial_x\psi_1(x) &= \pi (\psi_1^\dagger(x)\psi_1(x))^2\\
    -i\psi^\dagger_2(x)\partial_x\psi_2(x) &= \pi (\psi_2^\dagger(x)\psi_2(x))^2.
\end{align*}
As a result, following the substitution $\theta \to \beta\theta$, the long-range sine-Gordon Hamiltonian reading
\begin{align}
    H&=\int{\mathrm{d}x\left[\dfrac{1}{2}\dot{\theta}^2(x) + \dfrac{1}{2}\left(\theta'(x) - \dfrac{2\pi\delta}{\beta}\right)^2 - \dfrac{M^2}{\beta^2}\cos(\beta\theta(x)) \right.}\notag\\
    &+\left.\dfrac{1}{6}\int_{1}^{N/2}{\dfrac{\mathrm{d}u}{u}\left(\theta'(x)\right)\left(\theta'(x+u) + \theta'(x-u) \right)}\right]\notag\\
    &- \dfrac{4\pi\delta}{3\beta}\ln\left(\dfrac{N}{2}\right)\int{\mathrm{d}x \theta'(x)}\notag
\end{align}
transforms into a fermionic Hamiltonian:
\begin{align}
    H &= \int{\mathrm{d}x}\left[ic\psi^\dagger\sigma^z\partial_x \psi + g\psi^\dagger_1\psi_1\psi^\dagger_2 \psi_2 + M_0 \psi^\dagger\sigma^x\psi\right.\notag\\
    &\left.+\dfrac{(2\pi)^2}{6\beta^2}\int_1^{N/2}\dfrac{\mathrm{d}u}{u}\rho(x)(\rho(x+u) + \rho(x-u)) \right]\notag\\
    &-h'\int{\mathrm{d}x \rho(x)}\notag.
\end{align}
The speed of light $c$, the contact interaction $g$, the fermion mass $M_0$ and the chemical potential $h'$ follow the definitions introduced in the main text.

This Hamiltonian describes the behaviour of the system in the Weyl representation of fermionic fields. We can cast the model in Dirac representation via the transformation $\psi \to S \psi$, where the matrix $S$ reads
\begin{equation*}
    S = \dfrac{1}{\sqrt{2}}\begin{pmatrix} 1 & 1 \\
    1 &  -1        
    \end{pmatrix}.
\end{equation*}
According to this prescription, $S\sigma^x S = \sigma^z$ and $S \sigma^z S = \sigma^x$. The Hamiltonian can then be expressed as in Eq.~\eqref{Eq4}, namely as $H=H_0-h' \int \mathrm{d} x \rho(x) $, where 

\begin{align}
    H_0 &= \int\mathrm{d} x\left[-i c \overline{\psi} \gamma^1 \partial_x\psi + \dfrac{g}{4}(\overline{\psi}\gamma^\mu \psi)(\overline{\psi}\gamma_\mu \psi) + M_0 \overline{\psi}\psi \right]\notag\\
     &+\dfrac{(2\pi)^2}{6\beta^2}\int{\mathrm{d} x\int_{1}^{N/2}\dfrac{\mathrm{d}u}{u}\rho(x)\left(\rho(x + u) + \rho(x - u) \right)}\notag.
\end{align}
The Dirac gamma matrices here read then $\gamma^0 = \sigma^x$ and $\gamma^1 = i\sigma^y$.

\section{Discretization of the fields}
\label{appB}

The (1+1)-Thirring Hamiltonian is discretized following the prescription: we introduce a single discrete fermionic field whose value on even sites correspond to $\psi_1$ and to $\psi_2$ on odd sites, hence $\psi_1(x=n) \equiv \hat{f}_{2n}$ and $\psi_2(x=n) \equiv \hat{f}_{2n+1}$\,. Therefore, integrals transform into discreet sums as $\int{\mathrm{d}x} \to \sum_n$ and the Thirring model becomes

\begin{align}
    \hat{H} &= -\dfrac{ic}{2}\sum_n{(\hat{f}^\dagger_{n+1}\hat{f}_n - \hat{f}^\dagger_{n}\hat{f}_{n+1})} + \dfrac{g}{2}\sum_n{\hat{f}_{n+1}\hat{f}_{n+1}\hat{f}_n \hat{f}_n}\notag\\
    &+\sum_{n}((-1)^n M_0 - h')\hat{f}^\dagger_n \hat{f}_n + \dfrac{2\pi^2}{3\beta^2}\sum_n\sum_{\vert r\vert>1}{\dfrac{1}{\vert r \vert}\hat{f}^\dagger_n\hat{f}_n \hat{f}^\dagger_{n+r}\hat{f}_{n+r}}\, .\notag
\end{align}

Discretized fermionic systems can be mapped onto spin models by the means of a Jordan-Wigner transformation that identifies the local number of fermions to the orientation of a spin-$\frac{1}{2}$: namely $\vert 1\rangle \leftrightarrow \vert \uparrow\rangle$ and $\vert 0 \rangle \leftrightarrow \vert \downarrow\rangle$. As a result, one can deduce a simple relation betwwen the fermionic and the spin operators: $\hat{f}^\dagger_n \hat{f}_n = \hat{S}^z_n + \frac{1}{2}=\hat{\sigma}^z_n$. And to reconcile both the fermionic and $SU(2)$ spin algebra, the spin-fermion transformation yields $\hat{\sigma}^-_n=\exp\left[-i\pi\sum_{k<n}\hat{f}^\dagger_k \hat{f}_k\right]\hat{f}_n$ and $\hat{\sigma}^+_n=\hat{f}^\dagger_n\exp\left[-i\pi\sum_{k<n}\hat{f}^\dagger_k \hat{f}_k\right]$. The corresponding spin-Hamiltonian then takes the form of the XXZ model with site-dependent magnetic field and long-range Ising interactions of Eq.\ \eqref{XXZ}.

\section{Kac scaling and mean-field limit}

In the following, we discuss the derivation of the mean-field model by means of the rescaling of the spring stiffness. In one dimension, the energy contribution of Coulomb interactions scales as $N\ln N$. We apply the so called Kac scaling $K =K_0/\ln N$, with $K_0$ independent of $N$. This is equivalent to rescaling the charge or the interparticle distance $d_0$ \cite{Morigi:2004L}. Using this scaling then $\beta^2=\beta_0^2\sqrt{\ln N}$ where 
\begin{equation}
   \beta_0^2=\left(\frac{2\pi}{a}\right)^2 \sqrt{\frac{2\hbar^2}{3mK_0}} 
\end{equation}
is the rescaled Planck's constant. Consequently 
\begin{eqnarray*}
    c&=&\frac{\beta_0^2\sqrt{\ln N}}{8\pi }+\frac{2\pi}{\beta_0^2\sqrt{\ln N}}\\
    g&=& - \frac{\beta_0^2\sqrt{\ln N}}{4}+\frac{4\pi^2}{\beta_0^2\sqrt{\ln N}}
\end{eqnarray*}
At leading order in an expansion in $1/{\ln N}$ then 
\begin{eqnarray*}
    c&\sim&\frac{\beta_0^2}{8\pi}\sqrt{\ln N}\\
    g &\sim& -\frac{\beta_0^2}{4}\sqrt{\ln N}\,.
\end{eqnarray*}
Moreover, using Kac's scaling the soliton mass acquires a scaling with $\sqrt{\ln N}$ from the dependence on both the stiffness $K$ as well as from $\beta^2$:
\begin{eqnarray}
&&  M_{0}= \left(\frac{8\pi^2}{3\beta^2}\frac{\pi V_0}{K_0a^2}\right)\sqrt{\ln N}\,. 
\end{eqnarray}
With this prescription, the scaling of the uniform magnetic field becomes:
\begin{eqnarray}
&&  h'_N\approx \frac{8\pi^2}{3\beta_0^2}\delta\sqrt{\ln N}\,,
\end{eqnarray}
Finally, the Coulomb interaction term is now multiplied by the factor $2\pi^2/(3\beta_0^2\sqrt{\ln N})$. After collecting the common factor $\sqrt{\ln N}$ (which was introduced by the time rescaling $t\to \tau$), the XXZ Hamiltonian, Eq.\ \eqref{XXZ}, takes the form
\begin{eqnarray*}   
    \hat{H}_{\rm AFM} &=& - \frac{\beta_0^2}{8\pi}\sum_n{(\hat{\sigma}^+_n \hat{\sigma}^-_{n+1} + \hat{\sigma}^-_n \hat{\sigma}^+_{n+1})}\\
    &-&\frac{\beta_0^2}{8}\sum_n{\hat{\sigma}^z_n\hat{\sigma}^z_{n+1}}\notag\\
    &+& \frac{8\pi^2}{3\beta_0^2}\sum_n{\left((-1)^n\frac{\pi V_0}{K_0a^2}-\delta\right)\hat{\sigma}^z_n}\\
    &+& \dfrac{2\pi^2}{3\beta_0^2{\ln N}}\sum_n\sum_{\vert r\vert>1}{\dfrac{1}{\vert r \vert}\hat{\sigma}^z_n\hat{\sigma}^z_{n+r}}\, ,
    \label{XXZ:1}
\end{eqnarray*}
which is now extensive. With this rescaling, now for $\beta_0^2\gg 1$ the ground state is the one of the short range XXZ Hamiltonian while for $\beta_0^2\ll 1$ the Hamiltonian reduces to Eq.\ \eqref{Eq:Bak}. Note that the limit $N\to\infty$ does not commute with the expansion on $\beta_0$, a property characteristic of long-range interactions \cite{CAMPA200957,Morigi:2004,Defenu:2023}.\end{appendix}

\end{document}